\documentclass[10pt,journal,compsoc]{IEEEtran}

\ifCLASSOPTIONcompsoc
  \usepackage[nocompress]{cite}
\else
  \usepackage{cite}
\fi

\AtBeginDocument{%
  \providecommand\BibTeX{{%
    \normalfont B\kern-0.5em{\scshape i\kern-0.25em b}\kern-0.8em\TeX}}}
    
\usepackage{xcolor}
\usepackage[noend,ruled,linesnumbered]{algorithm2e}
\usepackage{algpseudocode}  
\usepackage{amsfonts,amsmath,bm} 
\usepackage{multirow,comment}
\usepackage{graphicx}

\newcommand{\dilname}{Dilithium }

\begin{document}

\title{A Unified Cryptoprocessor for Lattice-based Signature and Key-exchange}

\author{Aikata Aikata, Ahmet Can Mert, David Jacquemin, Amitabh Das,\\ Donald Matthews, Santosh Ghosh, Sujoy Sinha Roy 
\IEEEcompsocitemizethanks{\IEEEcompsocthanksitem Aikata, Ahmet Can Mert, David  Jacquemin, and Sujoy Sinha Roy are with the Institute of Applied Information Processing and Communications, Graz University of Technology, Graz, Austria.\\

E-mail:\\ \{aikata, ahmet.mert, david.jacquemin, sujoy.sinharoy\}@iaik.tugraz.at
\IEEEcompsocthanksitem Amitabh Das and Donald Matthews are with AMD, Austin, Texas, US.\\
E-mail: \{Donald.Matthews, Amitabh.Das\}@amd.com
\IEEEcompsocthanksitem Santosh Ghosh is with Intel Labs,  Intel Corporation, OR, US.\\
E-mail: santosh.ghosh@intel.com}

}

\IEEEtitleabstractindextext{
\begin{abstract}
We propose design methodologies for building a compact, unified and programmable cryptoprocessor architecture that computes post-quantum key agreement and digital signature. Synergies in the two types of cryptographic primitives are used to make the cryptoprocessor compact. 
As a case study, the cryptoprocessor architecture has been optimized targeting the signature scheme 'CRYSTALS-Dilithium' and the key encapsulation mechanism (KEM) 'Saber', both finalists in the NIST’s post-quantum cryptography standardization project. The programmable cryptoprocessor executes key generations, encapsulations, decapsulations, signature generations, and signature verifications for all the security levels of Dilithium and Saber. 
On a Xilinx Ultrascale+ FPGA, the proposed  cryptoprocessor consumes 18,406  LUTs,  9,323  FFs, 4 DSPs, and  24 BRAMs. It achieves 200 MHz clock frequency and finishes CCA-secure key-generation/encapsulation/decapsulation operations for LightSaber in 29.6/40.4/ 58.3$\mu$s; for Saber in 54.9/69.7/94.9$\mu$s; and for FireSaber in 87.6/108.0/139.4$\mu$s, respectively. It finishes key-generation/sign/verify operations for Dilithium-2 in 70.9/151.6/75.2$\mu$s; for Dilithium-3 in 114.7/237/127.6$\mu$s; and for Dilithium-5 in 194.2/342.1/228.9$\mu$s, respectively, for the best-case scenario. On UMC 65nm library for ASIC the latency is improved by a factor of two due to a 2$\times$ increase in clock frequency.
\end{abstract}
}

\maketitle

\IEEEdisplaynontitleabstractindextext
\IEEEpeerreviewmaketitle
\IEEEraisesectionheading{\section{Introduction}}

\IEEEPARstart{S}{hor’s} quantum algorithm \cite{shor} solves the integer factorization and discrete logarithm problems using quantum computers in polynomial time. These number theoretic problems are the foundations of the two most widely used  public-key cryptosystems, namely the RSA and Elliptic Curve cryptosystems. Hence, if a sufficiently powerful quantum computer is ever constructed, then the present-day public-key cryptographic schemes can be broken using Shor’s algorithm. Post-quantum cryptography (PQC) aims at developing new cryptographic protocols that will remain secure even after powerful quantum computers are built.

Acknowledging the fast progress~\cite{sycamore,chinese} made towards the development of quantum computers, different cybersecurity agencies have recommended gradual transitions toward post-quantum public-key encryption (PKE), key encapsulation mechanism (KEM), and  digital signature algorithm.  
For example, the Chinese Association for Cryptologic Research already concluded their post-quantum cryptography standardization `competition' with LAC~\cite{lac} as the winner key agreement scheme. Another standardization project, launched by the American National Institute of Standards and Technology (NIST) in 2016, is currently in its third round with four finalists in the PKE/KEM category, and three finalists in the digital signature category. NIST will be announcing the standardized PQC schemes in March 2022. 

As part of the transition phase, present-day RSA and elliptic curve-based PKE/KEM and signature schemes in various cryptographic applications will get replaced by their post-quantum counterparts. One of the most widely used applications is the Transport Layer Security (TLS) protocol, which ensures the security and privacy of communications over a computer network. TLS requires both a PKE/KEM for secure key exchange and a digital signature scheme for authentication. Therefore, a cryptoprocessor that runs the TLS protocol, will need to provide support for both PKE/KEM and signature algorithms. In a connected world, the majority of the cryptoprocessors will be running on resource-constrained platforms with strict silicon and memory budgets. 
Therefore, researching design methodologies for compact and unified implementations of post-quantum PKE/KEM and signature is required to make PQC deployable in real-life applications.
In this paper, we explore this direction and develop design methodologies for unifying lattice-based PKE/KEM and signature algorithms in a compact silicon area.

Various works exist in the literature (\cite{DBLP:journals/tches/RoyB20,gaj_fpt_19,hecompact,abdulgadirlightweight,mera_compact_20,imran2021design,lwrpro_21,zhou_dil_21,ricci_dil_21,DBLP:journals/iacr/LandSG21,gaj_dil_21,dilithium_2022,kyb1,falcon,ntru,abs,GhoshMKDGVS22}) that present optimized implementations of either a PKE/KEM or a signature scheme. While such works show how to implement a given PQC algorithm optimally, they do not take real-world applications (i.e., both PKE/KEM and signature) into consideration. There are only a very few cryptoprocessors~\cite{sapphire_2019,risqv_2020,fritzmann_masked_21,DangMG21} that can support more than one PQC protocol.~\cite{sapphire_2019} is a cryptoprocessor coupled with RISC-V processor implemented in ASIC for various lattice-based Round 2 candidate schemes in NIST's PQC standardization. It does not provide support for the latest finalists. In~\cite{risqv_2020,fritzmann_masked_21} the authors propose unified architectures for multiple PKE/KEM schemes only. They do not support any digital signature scheme and are therefore not ideal for the TLS or similar applications.  \textcolor{black}{ The authors in \cite{DangMG21} only present results for the three PKE/KEM schemes and do not include a digital signature scheme.} Thus, the compact and unified implementation of the two types of cryptographic primitives is a less explored problem.

We propose design methodologies for realizing a compact yet fast cryptoprocessor for performing both lattice-based signature and key-exchange operations. In this work, we focus on \emph{lattice-based} schemes because there are several PKE/KEM and signature schemes based on lattice problems. 
In the Status Report on the Second Round of the NIST PQC Standardization Process~\cite{nist_r2_report}, NIST mentioned that PKE/KEM-finalist CRYSTALS-Kyber and signature finalist CRYSTALS-Dilithium (which we refer to as 'Dilithium' for the rest of the paper) `have a common framework'. However, no such synergies have been indicated for the other lattice-based finalist candidates Saber, NTRU, or Falcon. Both CRYSTALS-Kyber and \dilname use Number Theoretic Transform (NTT) friendly parameter sets. Therefore, they have some obvious similarities. Interestingly, synergies in the other lattice-based schemes have not been explored yet. As a case study, in this paper, we choose the signature finalist \dilname~\cite{dilithium_nist_round3} and PKE/KEM finalist Saber~\cite{saber_nist_round3} for the unified cryptoprocessor. While \dilname is NTT-friendly, Saber is NTT-unfriendly. Therefore, they seem to have no obvious similarities, which makes our work interesting.
\newline

\hspace{-.5cm}\textbf{Our Contributions:}
Our goal is to design a unified and flexible cryptoprocessor architecture that can compute the lattice-based signature \dilname and PKE/KEM Saber with a balance between area and performance. 
We make the following contributions towards our goal.
\begin{enumerate}
    \item Polynomial multiplication is a time- and area-consuming operation in both schemes. 
    To realize a unified polynomial multiplier, we use the NTT method of polynomial multiplication for both \dilname and Saber. That particular algorithmic choice is made as \dilname~\cite{dilithium_nist_round3} makes NTT-based polynomial multiplication an integral part of the scheme.
    To use NTT-based polynomial multiplication in Saber, we show how to choose an appropriate prime for Saber's NTT. Next, we show that by using \dilname's prime, Saber could use the polynomial multiplier of \dilname readily at the cost of a negligible execution error probability. Finally, we design a unified NTT multiplier for \dilname and Saber.
    
    \item Both Dilithium and Saber make use of Keccak-based pseudo-random number generations and hash calculations. At the same time, the two schemes perform pre-and post-processing of the data at the input and output of the Keccak module differently. To make our cryptoprocessor compact without compromising speed, we implement an optimized wrapper around the Keccak module for performing scheme-specific pre/post-processing of data on the fly. That reduces both the area and the cycle counts significantly. 
    
    \item The building blocks that are specific to Dilithium or Saber, are optimized to reduce their memory access overheads and area. Although these scheme-specific blocks have a linear time complexity, we observe that their optimized implementations (for all the security parameters) in hardware require low-level bit and word manipulations. 
    
    \item Starting from the optimized building blocks, we construct a programmable instruction-set architecture (ISA). The ISA computes all the signature and KEM routines of \dilname and Saber and supports all the security levels of the two schemes.  
    
    \item The designed ISA offers the parallelism to execute several data-independent instructions concurrently so that the KEM and signature operations get faster. The data memory of the ISA is organized to enable concurrent reads/writes by the parallel instructions.
    With the parallel execution of instructions, we obtained significant reductions in the number of cycles for both Saber (e.g., 10\%, 13\%, and 15\% during decapsulation for LightSaber, Saber, and FireSaber respectively) and Dilithium (e.g., 20\%, 25\%, and 28\% during signature generation for Dilithium 2, 3, and 5 respectively).

\end{enumerate}

Although as a case study we took Dilithium and Saber, a similar methodology can be used for making unified cryptoprocessors for the other PKE/KEM and Signature schemes. The designed instruction-set architecture can also be generalized to support other algorithms, thus making it 'crypto-agile', which is a desired and a much-required aspect of modern-day cryptoprocessor design.

The paper is organized as follows. Sec.~\ref{sec:math_background} briefly describes the Dilithium and Saber protocols and their internal subroutines. Sec.~\ref{sec:syn} identifies the synergies in the two schemes. Next, in Sec.~\ref{sec:design_dec} the synergies are used to design the cryptoprocessor. The section describes the design methodologies in detail. Area and performance results are presented in Sec.~\ref{sec:results}. Discussions on extending the cryptoprocessor for supporting other schemes are presented in Sec.~\ref{sec:disc}. The final section draws the conclusions.

\section{Preliminaries}
\label{sec:math_background}
This section gives the specifications of Dilithium and Saber. Saber~\cite{saber_nist_round3} is an IND-CCA secure KEM and its security relies on the hardness of the Module Learning With Rounding (MLWR) problem. It has three variants: LightSaber, Saber, and FireSaber targeting different security levels. All of these variants use the same polynomial rings $R_q =\mathbb{Z}_q[x]/\langle x^{n} + 1 \rangle$ and $R_p =\mathbb{Z}_p[x]/\langle x^{n} + 1 \rangle$ with polynomial degree $n=256$ and the power-of-two moduli $q=2^{13}$ and $p=2^{10}$. The three variants use different module-dimensions and secret-distributions. Dilithium~\cite{dilithium_nist_round3} is a digital signature scheme and  its  security  is  based  on  the  computational hardness of the Module Learning With Errors (MLWE) and Module Short Integer Solution (MSIS) problems.  Depending on the size of the module $R_{q}^{k \times \ell}$ with $k, \ell > 1 $, Dilithium also comes with three variants, namely Dilithium-2, 3, and 5 for the NIST-specified security levels 2, 3, and 5 respectively~\cite{dilithium_nist_round3}.  All the three variants of Dilithium use the polynomial ring $R_q =\mathbb{Z}_q[x]/\langle x^{n} + 1 \rangle$ with $n=256$ and $q=2^{23}-2^{13}+1$, a prime modulus.

\subsection{Saber modules}
We briefly describe the internal routines of Saber. For a detailed description of them, readers may follow the original specification of Saber~\cite{saber_nist_round3}.
\begin{itemize}
    \item $\mathtt{gen()}$: It expands a uniform seed $\rho \in  \{0, 1\}^{256}$ using the Keccak-based expandable output function SHAKE-128 and generates the public matrix $\pmb{A} \in R_{q}^{l \times l}$.
    \item $\beta_\mu ()$: It samples a secret polynomial vector ($\pmb{s}$) from a binomial distribution with the parameter $\mu$.     
    \item Hash functions: Saber uses three hash functions: $\mathcal{F}()$, $\mathcal{H}()$ and $\mathcal{G}()$. The $\mathcal{F}()$ and $\mathcal{H}()$ are implemented using SHA3-256 while $\mathcal{G}()$ is implemented using SHA3-512. All hash functions are Keccak-based.
    \item Polynomial arithmetic: They include polynomial multiplication, polynomial addition/subtraction, coefficient-wise rounding using bit-shifting, equality checking of two polynomials, etc.
    \item $\mathtt{Verify}$ and $\mathtt{CMOV}$: $\mathtt{Verify}$ is used to perform ciphertext equality check during decapsulation. The result of Verify is stored in a flag register that is used by $\mathtt{CMOV}$ (constant-time move) to either copy the decrypted session key or a pseudo-random string at a specified location.
    \item $\mathtt{AddPack}$: 
    It performs the operation $(v + \textbf{h}_1 - 2^{\epsilon_p-1}m$ mod $p).(\epsilon_p - \epsilon_T)$ on a message bits $m$ using precomputed polynomial $v$ and constant $\textbf{h}_1$.
    \item $\mathtt{UnPack}$:  
    It performs the operation $(v + \textbf{h}_2 - 2^{\epsilon_p-\epsilon_T}c_m$ mod $p)(\epsilon_p - 1)$ on a ciphertext $c_m$ using precomputed polynomial $v$ and constant $\textbf{h}_2$.
    \item $\mathtt{AddRound}$: 
    It performs the operation $(((\textbf{A}^T + \textbf{s})$ mod $q).(\epsilon_q - \epsilon_p))$ using pubilic matrix vector $\textbf{A}$ and secret vector $\textbf{s}$.
\end{itemize}

\vspace{-1em}

\subsection{Dilithium modules}
We briefly describe the internal routines of Dilithium. For a detailed description of them, readers may follow the original specification of Dilithium~\cite{dilithium_nist_round3}.
\begin{itemize}
\item $\mathtt{ExpandA}()$: This function uses SHAKE-128 to generate the polynomials of the public matrix $\pmb{A} \in R_q^{k\times \ell}$ in parallel by expanding the common seed $\rho \in \{0,1\}^{256}$ along with different 16-bit nonce values. 
\item $\mathtt{ExpandS}()$: It is used to generate the secret polynomial vectors $\pmb{s}_1$ and  $\pmb{s}_2$ $\in S_\eta^{\ell} \times S_\eta^k$. For each polynomial the seed $\varsigma$ and a 16-bit nonce are fed to SHAKE-256 and the squeezed output is given to the rejection sampler for sampling the signed values in the range $\{-\eta,\eta\}$.
\item $\mathtt{Power2Round}_q()$:  This function takes an element $r = r_1\cdot 2^d  + r_0 $ and returns $r_0$ and $r_1$, where $r_0 = r\mod^{\pm} 2^d$ and $r_1 = (r-r_0)/2^d$.
\item $\mathtt{HighBits}_q() \textnormal{ and } \mathtt{LowBits}_q()$: Let $\alpha$ be a divisor of $q-1$. The function $\mathtt{Decompose}_q()$ is defined in the same way as $\mathtt{Power2Round}()$ with $\alpha$  replacing $2^d$ in $\mathtt{Power2Round}()$. 
\item $\mathtt{MakeHint}_q()$/$\mathtt{UseHint}_q()$: $\mathtt{MakeHint}$ uses $\mathtt{Decompose}_q()$ to produce a hint $\pmb{h}$. $\mathtt{UseHint}$ uses the hint $\pmb{h}$ produced by $\mathtt{MakeHint}_q()$ to recover the high-bits.
\item $\mathtt{CRH}()$: This is a collision-resistant hash function which utilizes 384 bits of the output of SHAKE-256. 
\item $\mathtt{SampleInBall}()$: It fills a polynomial with only $\tau$ coefficients set to $+1$ or $-1$ and the remaining coefficients as 0. 
\item $\mathtt{ExpandMask}()$: This function expands ($\acute{\rho} \parallel \kappa$) string to generate a polynomial vector. The SHAKE output is broken into a sequence of positive integers in the range $[0, 2\gamma_1-1]$ and these are processed using a rejection sampling. 
\item Polynomial Arithmetic and $\mathtt{NTT}()$: Polynomial multiplications are performed using the NTT method. 

\end{itemize}
The signing operation generates a potential signature and checks a set of constraints on the generated signature. If satisfied, a valid signature is produced as the output; otherwise, the loop continues with generating another potential signature.

\section{Synergies and Design Decisions}\label{sec:syn}

To design a compact and unified cryptoprocessor that supports multiple schemes, the most important step is to identify synergies in the cryptographic schemes. The synergies will make resource-sharing possible and therefore reduce area and memory overheads. 

In this work, we aim for a unified architecture for Dilithium and Saber schemes as a case study. Therefore, we will focus on synergies between these two schemes and the corresponding design decisions. In this section, we identify the most important synergies in Dilithium and Saber.

Both Dilithium and Saber are based on module lattices and therefore they share several structural similarities. For example, both schemes operate on matrices and vectors of polynomials where the polynomials are always of 256 coefficients. Hence, the underlying polynomial arithmetic operators are common to Dilithium and Saber. Furthermore, both schemes use Keccak-based hash functions and pseudo-random number generators. Interestingly, the accumulated time spent on polynomial multiplications, pseudo-random number generations, and hash calculations in the two protocols is around 90\% of the overall protocol execution time~\cite{cortexm4}. In summary, we observe that the most time-critical primitives in Dilithium and Saber have similarities.  

\subsection{Polynomial multiplication, pseudo-random number generation, and hash computation}

We have two options for implementing polynomial multiplications in Dilithium and Saber. The first option is to instantiate an NTT-based multiplier for Dilithium (which uses a prime modulus) and a schoolbook or Toom-Cook or Karatsuba multiplier for Saber (which uses power-of-two moduli) following~\cite{DBLP:journals/tches/RoyB20} so that both schemes can be executed at their optimal speeds. One big disadvantage of this approach is that the architecture results in a large area requirement due to the presence of scheme-specific multipliers. Furthermore, a large area requirement could potentially slow down the clock frequency of the implementation due to the increased routing and placement complexities. The other option is to instantiate a common polynomial multiplier for both Dilithium and Saber. In this case, the common polynomial multiplier must be NTT-based as the specification of Dilithium~\cite{dilithium_2022} makes the use of NTT-based multiplication an integral part of the protocol. Saber~\cite{saber_nist_round3} could use any type of polynomial multiplication method, including the NTT-based one.  

Both Dilithium and Saber use the Keccak-based hash function SHA3 and pseudo-random number generator SHAKE. Therefore, both schemes could use a common Keccak core along with an appropriate wrapper around the core for realizing different SHA3 and SHAKE functionalities.

\subsubsection{Prime selection for Saber}
\color{black}
To use the NTT-based polynomial multiplication in Saber, a sufficiently large NTT-friendly prime $q'$ is needed to ensure the correctness of the multiplication~\cite{DBLP:journals/tches/ChungHKSSY21}. The polynomial multiplication will be erroneous if a true modular reduction by $q'$ takes place during the internal computation steps.  
We have two main options for choosing the prime $q'$. The first option is to choose a sufficiently large $q'$ to ensure no true modular ever happens. The other option is to use a relatively smaller prime that offers some computational advantages over the first option while keeping the error probability negligible. We describe both options in the following part of this subsection.

\noindent\textbf{Prime selection for error-free multiplication in Saber:} 
To select a prime for Saber which will work for all security levels of Saber without any error, we consider the maximum value that a polynomial coefficient can have after the polynomial multiplication operation. 
In Saber, one of the input polynomials of a polynomial multiplication operation is always a secret key polynomial which has small coefficients in the range $[-\frac{\mu}{2},\frac{\mu}{2}]$ where $\mu$ is 10, 8, 6 for LightSaber, Saber, FireSaber, respectively~\cite{saber_nist_round3}.
Similarly, the second polynomial operand of the polynomial multiplication operation in the Saber scheme is always a polynomial with coefficients in the range $[0,q-1]$ where $q=2^{13}$. 
If we remap the coefficients in $[0,q-1]$ to respective coefficients in $[-\frac{q}{2},\frac{q}{2}-1]$, then the maximum value that a coefficient of the polynomial multiplication result can have is $\frac{\mu}{2}\cdot \frac{q}{2} \cdot n = 5 \cdot 2^{12} \cdot 2^8$. Similarly, the minimum value that a coefficient of the result can have is $-5 \cdot 2^{12} \cdot 2^8$. Therefore, a 24-bit prime $q'_{24}$ can be used in Saber as it will be larger than $2 \cdot 5 \cdot 2^{12} \cdot 2^8$ and therefore no true modular reductions will ever take place during a polynomial multiplication. 

In Saber, polynomial matrices and vectors are multiplied and therefore several polynomial multiplication results are accumulated. These accumulations can be performed directly in the NTT domain to avoid unnecessary inverse NTT transforms. Taking the accumulation into consideration, the maximum and the minimum values that a coefficient of the accumulated result can have will be larger than $\frac{\mu}{2}\cdot \frac{q}{2} \cdot n$ or lower than $-\frac{\mu}{2}\cdot \frac{q}{2} \cdot n$, respectively. The exact maximum value will depend on the dimensions of the matrices and vectors.
The public polynomial matrix of Saber has the dimension $l\times l$ where $l$ is 2, 3, and 4 for LightSaber, Saber, and FireSaber respectively~\cite{saber_nist_round3}. The vectors are of size $l$ polynomials.
Hence, the maximum value that a coefficient can take after a matrix-vector multiplication is $l \cdot \frac{\mu}{2}\cdot \frac{q}{2} \cdot n$, which will be at most $3 \cdot 4 \cdot 2^{12} \cdot 2^8$
 when Saber or FireSaber is used.
Similarly, the minimum value that a coefficient can take after a matrix-vector multiplication is $-3 \cdot 4 \cdot 2^{12} \cdot 2^8$.
Therefore, $q'$ is required to be at least 25-bit to ensure the zero error probability for the matrix-vector multiplication in Saber.
We selected the 25-bit prime $q'_{25}=2^{25}-2^{14}+1$ for implementing the error-free matrix-vector multiplication of Saber. The sparse structure of the prime enables a fast and lightweight modular reduction circuit. 

\noindent\textbf{Prime selection for ensuring a negligible error probability:} 
In the previous subsection, we saw that a 25-bit prime is required when we consider the worst-case scenario. In practice, the probability of having a polynomial coefficient close to the worst-case bound is negligible as the coefficients of the secret polynomials are binomially distributed in the range $[-\frac{\mu}{2},\frac{\mu}{2}]$ and the coefficients of the public polynomials are uniformly distributed in the range $[0,q-1]$. 
One attractive choice is the 23-bit prime modulus of Dilithium. With this special prime, Saber can trivially use the NTT-based polynomial multiplier of Dilithium for computing its own polynomial multiplications. At the same time, we must be sure that the smaller prime does not cause a non-negligible error probability.

Now we describe a method for estimating the probability of getting an error when a prime smaller than 25-bit is used in Saber's NTT. Note that, an error will happen when any resultant coefficient after a matrix-vector multiplication satisfies $|\text{coeff.}| > q'/2$ causing a true modular reduction by $q'$.
To keep the analysis simple, we assume that the secret coefficients are also uniformly distributed in $[-\frac{\mu}{2},\frac{\mu}{2}]$. When we perform a polynomial multiplication between a secret polynomial and a public polynomial, both of which follow uniform distributions, Irwin-Hall states that the coefficients of the resultant polynomial will follow a normal distribution~\cite{johnson1995continuous}. \textcolor{black}{Note that a binomially distributed polynomial is significantly sparse compared to a uniformly distributed polynomial in the same range. Hence, the assumption that the secret coefficients follows uniform distribution enables a more conservative approximation (i.e., an upper bound) of the computation error probability.}

\begin{figure}[t]
    \centering
    \includegraphics[width=0.44\textwidth]{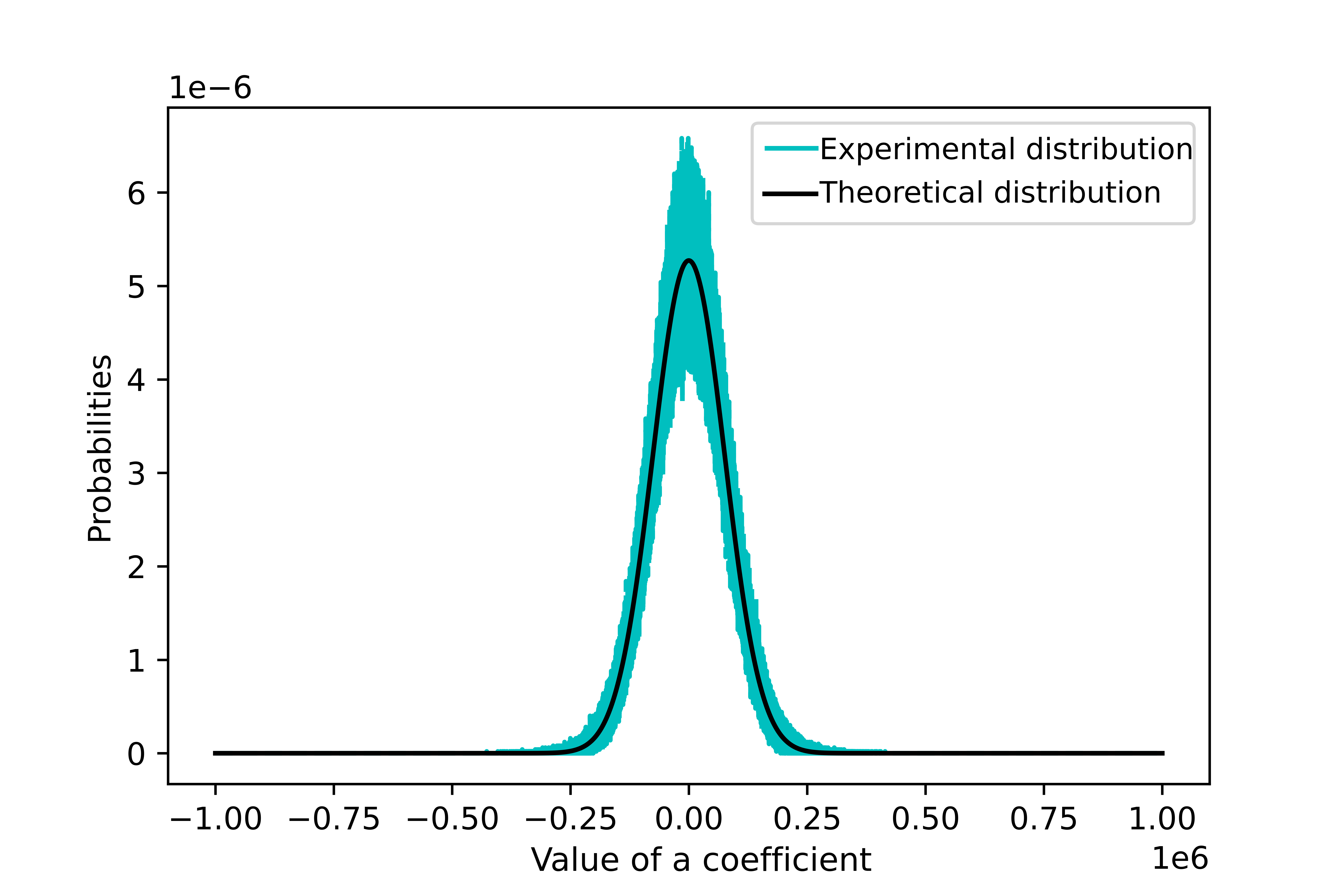}
    \vspace{-1em}
    \caption{Distribution of a coefficient after a multiplication between a secret and a public polynomial. The secret and public coefficients are in the range $[-\frac{\mu}{2}, \frac{\mu}{2}]$ and $[0, q - 1]$, respectively.}
    \label{fig:plot}
\end{figure}

We experimentally verified the above statement. We used a python script to generate pairs of polynomials where the coefficients of the first polynomial are uniformly random in the range $[-\frac{\mu}{2},\frac{\mu}{2}]$. \textcolor{black}{For the coefficients of the second polynomial, we experimented on both $[-q/2, q/2-1]$ and $[0, q-1]$, and selected $[0, q-1]$ to get the more conservative measurement}. Then the two polynomials within a pair are multiplied. The experimentation was performed for millions of independent pairs. The scattered cyan plot in Fig.~\ref{fig:plot} shows the distribution of the resultant coefficients from our experimentation. We observe that the experimental distribution closely resembles a normal distribution, and therefore the Irwin-Hall statement is found to be valid in this polynomial multiplication scenario. From the experimental data, the mean and the standard deviation were calculated to plot the theoretical normal distribution (in black color) in Fig.~\ref{fig:plot}. 

When such $l$ independent polynomial multiplication results are accumulated, the resultant coefficients will follow a normal distribution with the mean and standard deviation increasing by $l$ and $\sqrt{l}$ times respectively. In Table~\ref{tab:Results} the distribution parameters are presented for all the three security levels of Saber.

Next, we used the cumulative distribution function of the normal distribution to estimate \color{black} the upper bound for \color{black} the error probability. Table~\ref{tab:Results} shows the error calculation for a 23-bit prime ($q'_{23}=2^{23}-2^{13}+1$) and a 24-bit prime ($q'_{24}=2^{24}-2^{14}+1$). We can see that \color{black} the upper bound \color{black} is extremely low for both the primes. Therefore, the use of Dilithium's 23-bit prime in Saber's NTT-based polynomial multiplication causes negligible error possibilities for all three security levels of Saber.

{The authors in \cite{duplicate} 
also provide an error probability estimate, when Dilithium's prime is used for Saber. They used combinatorics to count all the possible combinations of a polynomial pair that can lead to an error after a polynomial multiplication. Their approach is specific to a \emph{single} polynomial multiplication as they do not take the accumulation of polynomial multiplication results into consideration for calculating the error probabilities.}

\vspace{-1em}
\begin{table}[t]
\centering
\caption{\label{tab:widgets}Distribution parameters and error probabilities}
\vspace{-1em}
\begin{tabular}{l|c|c|c}
\hline
        & \textbf{LightSaber} & \textbf{Saber}  & \textbf{FireSaber} \\
        \hline
Mean & 115.72 & 48.06 & -29.32\\
Std. dev. & 119708.81 & 97825.69 & 75672.40\\ \hline
Accumulated mean & 231.45 & 144.18 & -117.28\\
Accumulated std. dev. & 169293.83 & 169439.06 & 151344.81\\ \hline
P($|\text{coeff.}|>q'_{24}/2$) & $2^{-1774}$ & $2^{-1837}$ & $2^{-2219}$\\
P($|\text{coeff.}|>q'_{23}/2$) & $2^{-449}$ & $2^{-448}$ & $2^{-558}$\\
\hline
\end{tabular}
\label{tab:Results}
\vspace{-1em}
\end{table}

\subsection{Remaining scheme-specific building blocks} \label{sec:rem_bb}

The remaining building blocks in the two schemes do not share many similarities and they mostly perform simpler operations (i.e., addition, packing) of linear time complexity compared to the polynomial multiplication and pseudo-random number generation operations. 
To further reduce the area consumption, an option could be to resource-share a common set of arithmetic circuits (e.g., addition and subtraction) with algorithm-specific finite state machines for generating the control signals. This design decision might decrease the area and at the same time might make the overall design complex and serial instead of parallel. 
Therefore, to make the design simple and easily configurable, we decide to keep the scheme-specific blocks separate in the implementation. However, we do perform optimizations within these blocks. For example, we combine all the different packing and unpacking methods required by Dilithium to make a unified pack/unpack unit that consumes much less area when compared to separate implementations.

\section{Optimized hardware architecture}\label{sec:design_dec}
In this section, we describe how we design a compact and unified architecture for the lattice-based signature scheme Dilithium and PKE/KEM Saber. Our target is to support all the subroutines and all the security levels of Dilithium and Saber in the same cryptoprocessor. Therefore, we choose the instruction-set architecture (ISA) framework for developing our unified cryptoprocessor. In this framework, the primitive building blocks are `instructions' that are called in an appropriate sequence during the execution of a cryptographic protocol. Therefore, programmability or flexibility is inherent to the ISA framework, which is a desirable feature for supporting multiple cryptographic schemes.

\begin{figure}
    \centering
    \includegraphics[width=0.36\textwidth]{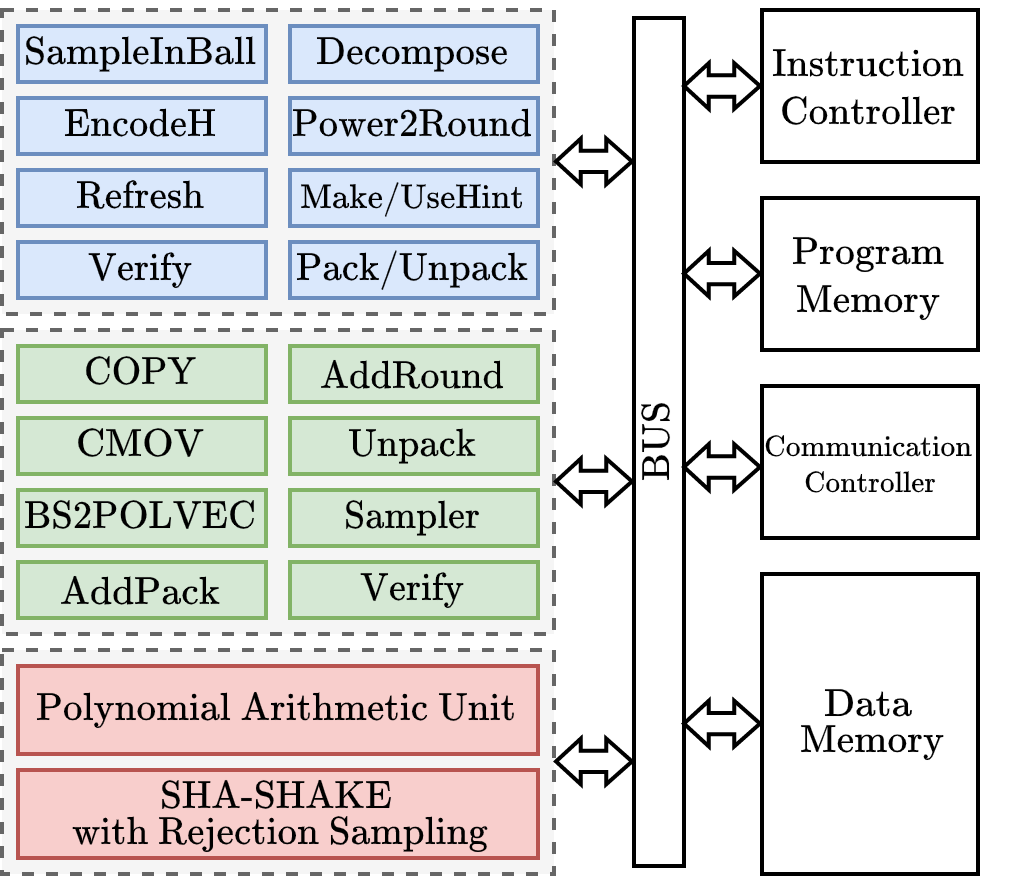}
    \vspace{-1em}
    \caption{The high-level design of the cryptoprocessor with the Saber, Dilithium and common modules are in green, blue and red, respectively.} 
    \label{fig:hwarch}
\end{figure}

The high-level block diagram of the proposed unified cryptoprocessor is shown in Fig.~\ref{fig:hwarch}.
The unified cryptoprocessor for Dilithium and Saber has a common NTT-based polynomial multiplier, a common Keccak-core (with a wrapper around it), and all the scheme-specific building blocks. The first two are the most expensive in terms of both computation time and area requirements, and thus they must be well optimized to make our unified cryptoprocessor compact and efficient. 

\subsection{NTT-based unified polynomial multiplier}\label{sec:ntt_mul}

This section describes the design decisions we make for implementing the NTT-based polynomial multiplier architecture for Dilithium and Saber.

\setlength{\textfloatsep}{0.15cm}
\setlength{\floatsep}{0.15cm}

\begin{figure}[t]
    \centering
    \includegraphics[width=0.45\textwidth]{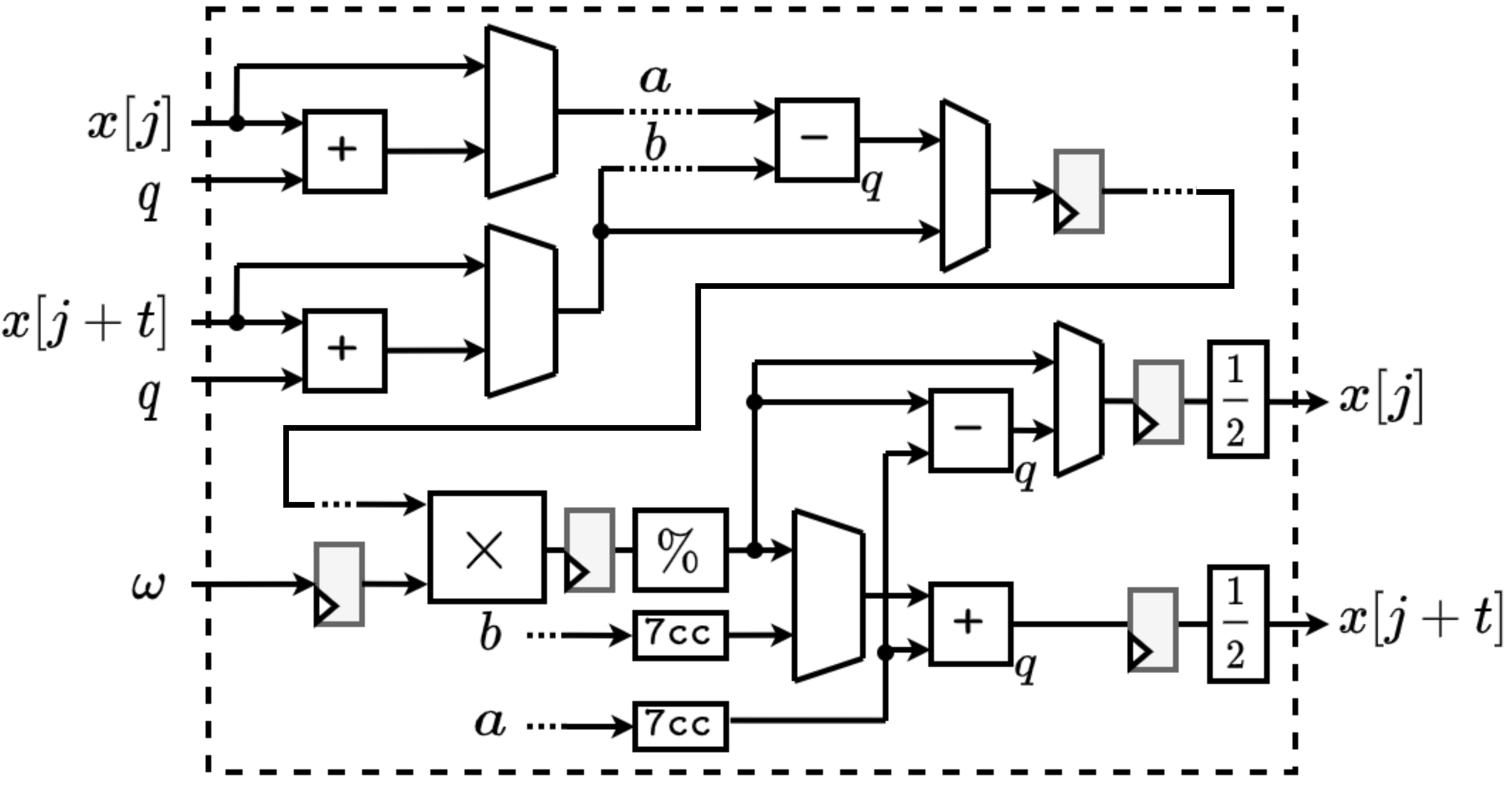}
    \vspace{-1em}
    \caption{Internal architecture of the unified butterfly unit. \textcolor{black}{The dotted routing lines for signals $a$ and $b$ are used to represent long interconnects} }
    \label{fig:bfu}
\end{figure}

Following the official reference code of Dilithium, we use the Cooley-Tukey (CT) 
and Gentleman-Sande (GS)  butterfly configurations for the NTT and inverse NTT (INTT) respectively. Fig.~\ref{fig:bfu} shows the internal blocks of the unified butterfly core. The proposed unified butterfly core can work with both CT and GS butterfly configurations. Both butterfly configurations perform operations using unsigned arithmetic in a unified butterfly core. The circuits are all pipelined to achieve high clock frequency. 

\noindent\textbf{Efficient modular reduction unit:} 
If Dilithium's prime is used for Saber's NTT, then just having an efficient modular reduction circuit for the 23-bit prime will be sufficient. Otherwise, if Saber uses the 24-bit or 25-bit primes (that are different from Dilithium's prime) then a resource-shared modular reduction circuit will be required. As the primes have a similar pseudo-Mersenne structure, designing and implementing a unified modular reduction unit will have a very small area overhead compared to a single prime-specific reduction unit. We followed the add-shift-based modular reduction method~\cite{DBLP:conf/date/YamanMOS21} and used a similar fast modular reduction technique for a unified modular reduction unit. Both Dilithium and Saber primes have the form of $q = 2^{x} - 2^{y} + 1$ which allows efficient modular reduction operation by using the property $  2^{x} \equiv 2^{y} - 1 \pmod{q}$ recursively.

When the Dilithium prime is used for both schemes, a modular reduction unit for $2^{23} - 2^{13} + 1$ is implemented and used for both schemes. The property $2^{23} \equiv 2^{13} - 1 \pmod{q}$ is used recursively (i.e., $c[45:23]\cdot2^{23} + c[22:0] = c[45:23]\cdot(2^{13} - 1) + c[22:0]$) to generate six partial results which are added efficiently using a carry save adder tree. Finally, a correction is performed to bring the result to the range $[0,q-1]$. 

\begin{figure}[!t]
    \centering
    \includegraphics[width=0.4\textwidth]{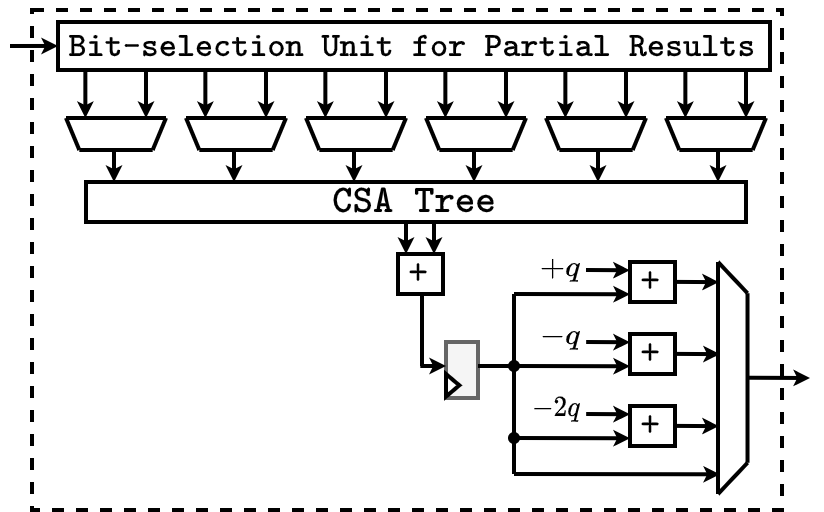}
    \vspace{-1em}
    \caption{Unified modular reduction unit}
    \label{fig:unified_mod_red}
\end{figure}

When Saber uses 25-bit (or 24-bit) prime, a unified modular reduction unit for $2^{23} - 2^{13} + 1$ and $2^{25} - 2^{14} + 1$ (or $2^{24} - 2^{14} + 1$) is implemented. Since both primes have similar structures, six partial results are generated for each reduction. Then, partial results are selected based on the scheme in use and the same steps are followed as modular reduction unit for the Dilithium prime. Fig.~\ref{fig:unified_mod_red} shows the high-level diagram of unified modular reduction unit.

\begin{figure}[!t]
    \centering
    \includegraphics[width=0.45\textwidth]{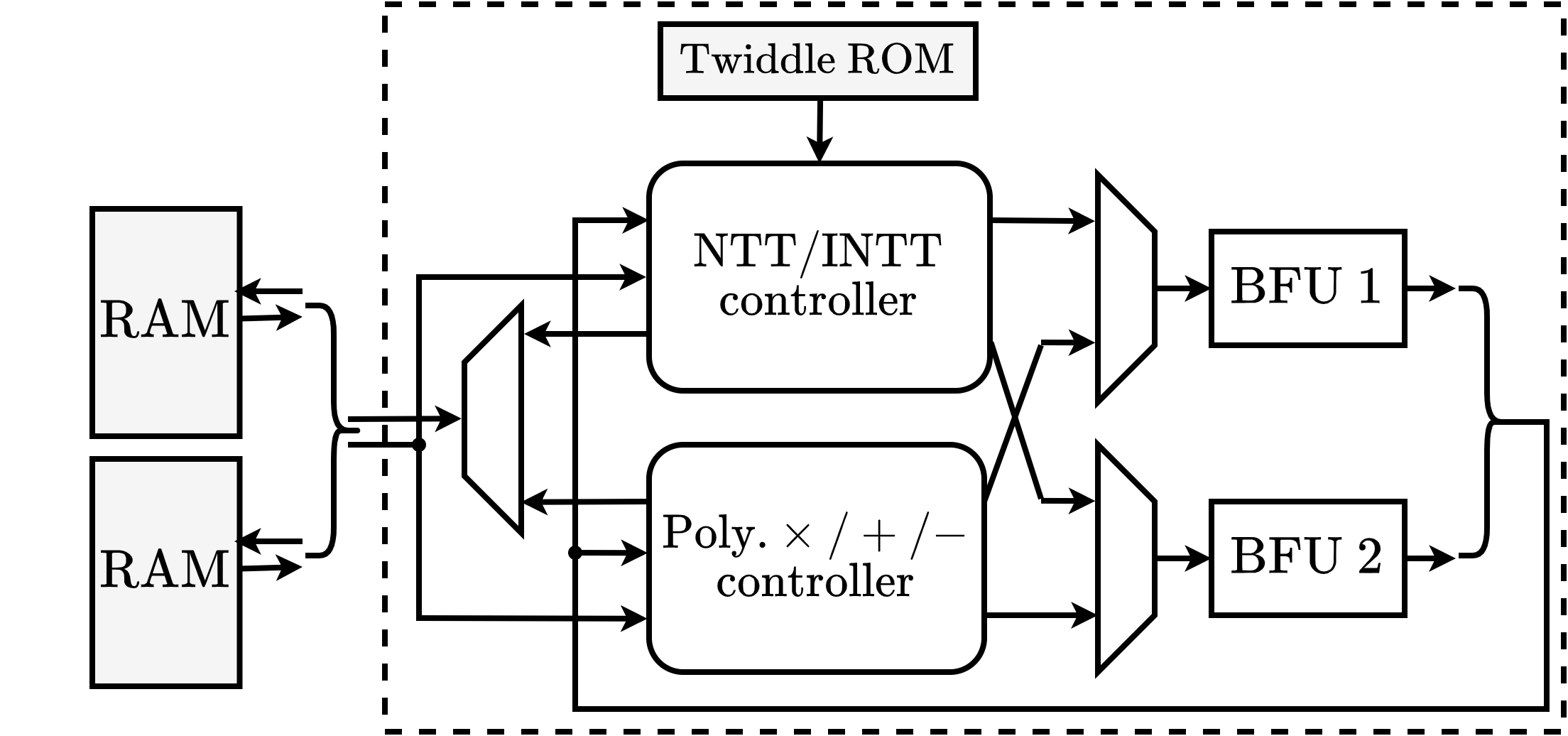}
    \vspace{-1em}
    \color{black}
    \caption{High-level architecture of polynomial arithmetic unit }
    \color{black}
    \label{fig:ntt_detailed}
\end{figure}

\noindent\textbf{Memory arrangement:} 
As one butterfly core consumes two coefficients and simultaneously produces two coefficients every cycle, we always keep two coefficients in a single memory-word following~\cite{roy_compact_lwe13}. This enables reading/writing two coefficients by just one memory-read/write. Our NTT unit has two such butterfly cores in parallel to reduce the cycle count of NTT. To feed the two butterfly cores, we spread the coefficients into two BRAM sets. This spreading is necessary as one BRAM-set could feed only one butterfly core due to the limitations in the number of read/write ports. In this way, a polynomial of 256 coefficients occupies a total of 128 memory words of which 64 are in the first BRAM set and the remaining 64 are in the other BRAM set. Note that generally, the efficient implementations propose that the coefficients which are to be processed are stored together, and this is maintained throughout the NTT/INTT iterations. However, since the coefficients are generated sequentially we plan to utilize them in the same way instead of preprocessing them to use the strategies existing in the literature.

\begin{figure}[b]
    \centering
    \includegraphics[width=0.48\textwidth]{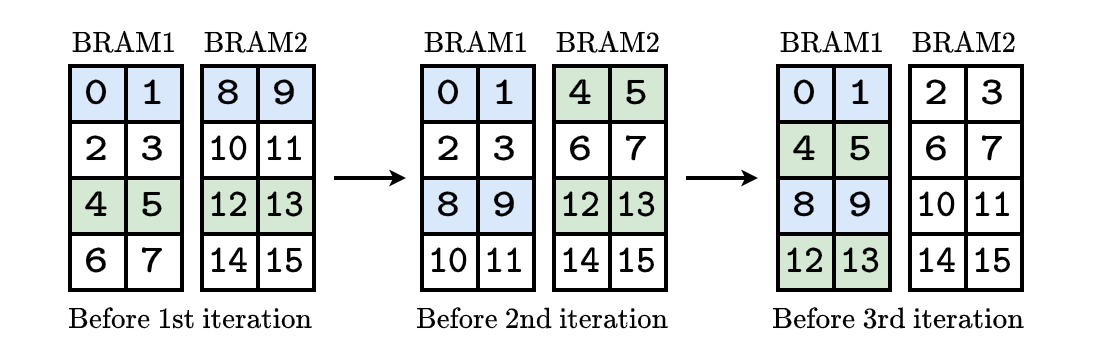}
    \vspace{-1em}
    \caption{Coefficients in memory for 3 iterations of NTT for $n=16$}
    \label{fig:rd_wt}
\end{figure}
Fig.~\ref{fig:rd_wt} shows the arrangement of coefficients in memory words during NTT loop-iterations using a toy example. During the NTT loops, the newly generated coefficients are written back in the BRAMs such that during the next iteration of the NTT loop, the required coefficients for each butterfly can be read as a pair from the memory. In the first iteration, the coefficients (shown in blue color in Fig.~\ref{fig:rd_wt}) zero-and-eight are input to the first butterfly unit and the coefficients one-and-nine are input to the second butterfly unit. After the first iteration, we want the processed coefficients eight-and-nine to be stored in BRAM1, at the address where currently coefficients four-and-five are stored. This will simplify the coefficient read pattern during the next iteration of the loop. 
Because coefficients four-and-five have not been processed yet, we can not write new values to their memory location. 
We solve this problem by changing the order in which the coefficients are processed. 
Hence, we read the coefficients four-and-five, and similarly twelve-and-thirteen immediately after reading the coefficients zero-and-one, and  eight-and-nine respectively. 

We add pipeline stages in between to avoid conflicts when writing coefficients eight-and-nine while reading coefficients four-and-five. Thus simplifying the control logic of the NTT and avoiding any read-write conflict. One NTT or INTT operations take 512 clock cycles only.  \textcolor{black}{ A high-level view of the complete polynomial arithmetic unit architecture is shown in Fig.~\ref{fig:ntt_detailed}. For simplicity, we have designed two separate controllers for NTT/INTT and remaining arithmetic operations. }

\noindent\textbf{Post-processing elimination after INTT operation:}
At the end of the INTT operation, the resulting coefficients are scaled by $1/n$, which requires extra $n$ modular multiplications. In our design, this extra scaling is removed by performing division by two in $\pmod{q}$ for each butterfly operation output in Gentleman-Sande configuration during INTT operation. The division by two in $\pmod{q}$ can be performed using only add and shift operations as shown in Eqn.~\ref{eqn:divby2}~\cite{DBLP:conf/date/YamanMOS21}. This way both the NTT and INTT implementations in our architecture are of the same cost and require no post-processing. This technique reduces the cost of INTT operation by 20\% at the expense of slight increase in the hardware logic.

\begin{equation}\label{eqn:divby2}
    \frac{x}{2} \pmod{q} = (x  \gg 1) + (x~\&~\texttt{0x1}) \times \frac{(q+1)}{2}
\end{equation}
\vspace{-2em}

\subsection{SHA3-256/512 and SHAKE-128/256 unit}

Many PQC schemes such as Dilithium and Saber use Keccak-based operations such as hash function SHA3 and pseudo-random number generator SHAKE.
For implementing the Keccak-based hash and expandable output functions, we instantiate a single high-speed Keccak core in the proposed cryptoprocessor architecture. Implementation of the Keccak core is similar to the high-speed Keccak core available on the website of Keccak-team~\cite{keccak_team}. We use a wrapper module around the Keccak core to perform parsing of input and output data bits.  Additionally, the state buffer has been changed so that the pseudo-random polynomial coefficients can be generated in scheme-specific optimal representations and then stored immediately in the memory of the cryptoprocessor. This strategy helps reduce the overall cycle counts for both Dilithium and Saber.

The type of parsing varies for different operation modes. For example, the public polynomials of Saber are generated by directly picking 13-bit coefficients from SHAKE-128 output. Whereas, the secret polynomials are generated by passing the SHAKE-128 output to a binomial sampler. Similarly, Dilithium's public and secret polynomials are generated using different types of rejection samplers. To explain the wrapper, we take Saber's public polynomial generation as a case study and later discuss how we use a similar methodology for different samplers.  

\noindent\textbf{Generating Saber's public polynomials:} Saber's public polynomials, generated using SHAKE-128, have a 13-bit coefficient size. Before these polynomials are multiplied, they are converted into the NTT representation in our unified cryptoprocessor. As described in Sec.~\ref{sec:ntt_mul}, the NTT unit requires its operand data to be present in `two coefficients per BRAM word' format, for reading and writing the coefficients efficiently. One option for processing the public polynomials will be to generate a continuous bitstream in 64-bit words (which is the default output format of Keccak), then store the words in BRAMs, and later parse them into 13-bit coefficients using separate parser hardware. This approach is sequential by nature and results in a bloated cycle count and area consumption due to required parsing buffers. To avoid such a redundant memory read/write step, we modify the output buffer of Keccak to directly produce a pair of 13-bit coefficients during the generation of the public matrix $\pmb{A}$. However, this strategy requires a book-keeping mechanism as the output length of a SHAKE-128 squeeze operation is 1,344 bits which is not a multiple of 13. Therefore, after each squeeze of SHAKE-128, there will be leftover bits that must be prepended to the output string generated by the next SHAKE-128 squeeze operation.  We observe that during the generation of $\pmb{A}$ in Saber, the number of leftover bits is always an even number in $[0, 24]$. We use this observation to simplify the implementation of the Keccak-output buffer. 

The prepending of the leftover bits to a newly generated SHAKE-128 squeeze output requires shifting and filling of the buffer bits. As the size of the Keccak output buffer (when operated as SHAKE-128) is 1,344 bits, which is quite large, we investigated efficient implementation techniques that reduce the area-overhead without affecting the cycle count. The naive method is to implement a simple multiplexer that updates the output buffer with 1,344 bits of the Keccak state and the leftover bits. But since there can be 13 (even numbers in [0, 24]) such possibilities we will require a 13-to-1 multiplexer for assigning to a buffer of size 1,368 (=1344+24) bits. With this implementation option, there are 13 shift possibilities and as a consequence, the multiplexing overhead is $\approx$8000 LUTs, which is large. We aim to make a very efficient and lightweight design on hardware, therefore we need a much better solution.

\begin{figure}[t]
    \centering
    \includegraphics[width=0.48\textwidth]{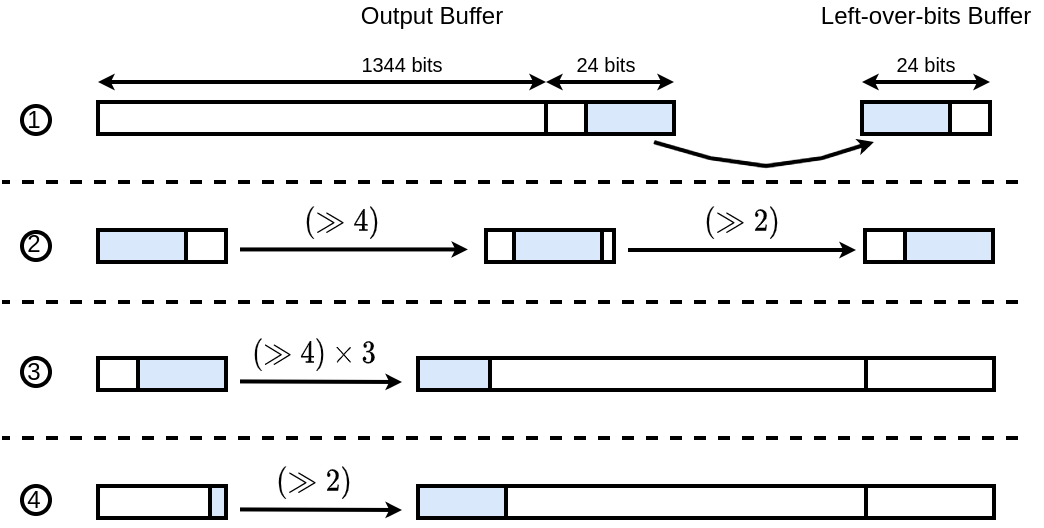}
    \vspace{-1em}
    \caption{Example of bookkeeping with 18 remaining bits}
    \label{fig:buff}
\end{figure}

We proposed an efficient method for handling the remaining bits using a small `left-over-bits buffer'. After the Keccak squeeze is done, we write the remaining bits to the left-over-bits buffer. To avoid using a multiplexer for deciding on the number of remaining bits we need to pick, we just write the last 24 bits, which is the maximum possible number of remaining bits. To put these remaining bits in the  Keccak output buffer, we first need to align the bits in the left-over-bits buffer to the left. In order to reduce the hardware cost, we decide to use only 'shift by two' or 'shift by four' operations for this alignment. Finally, the content of this left-over-bits buffer is concatenated at the beginning of the Keccak output buffer, using only 'shift by two' or 'shift by four' operations. An example for handling 18 leftover bits using the proposed left-over-bits buffer method is illustrated in Fig.~\ref{fig:buff}. Since we run Keccak in parallel with NTT, the extra cycle count for this bookkeeping does not account for an increase in the total cycle count.

\begin{figure}[t]
    \centering
    \includegraphics[width=0.35\textwidth]{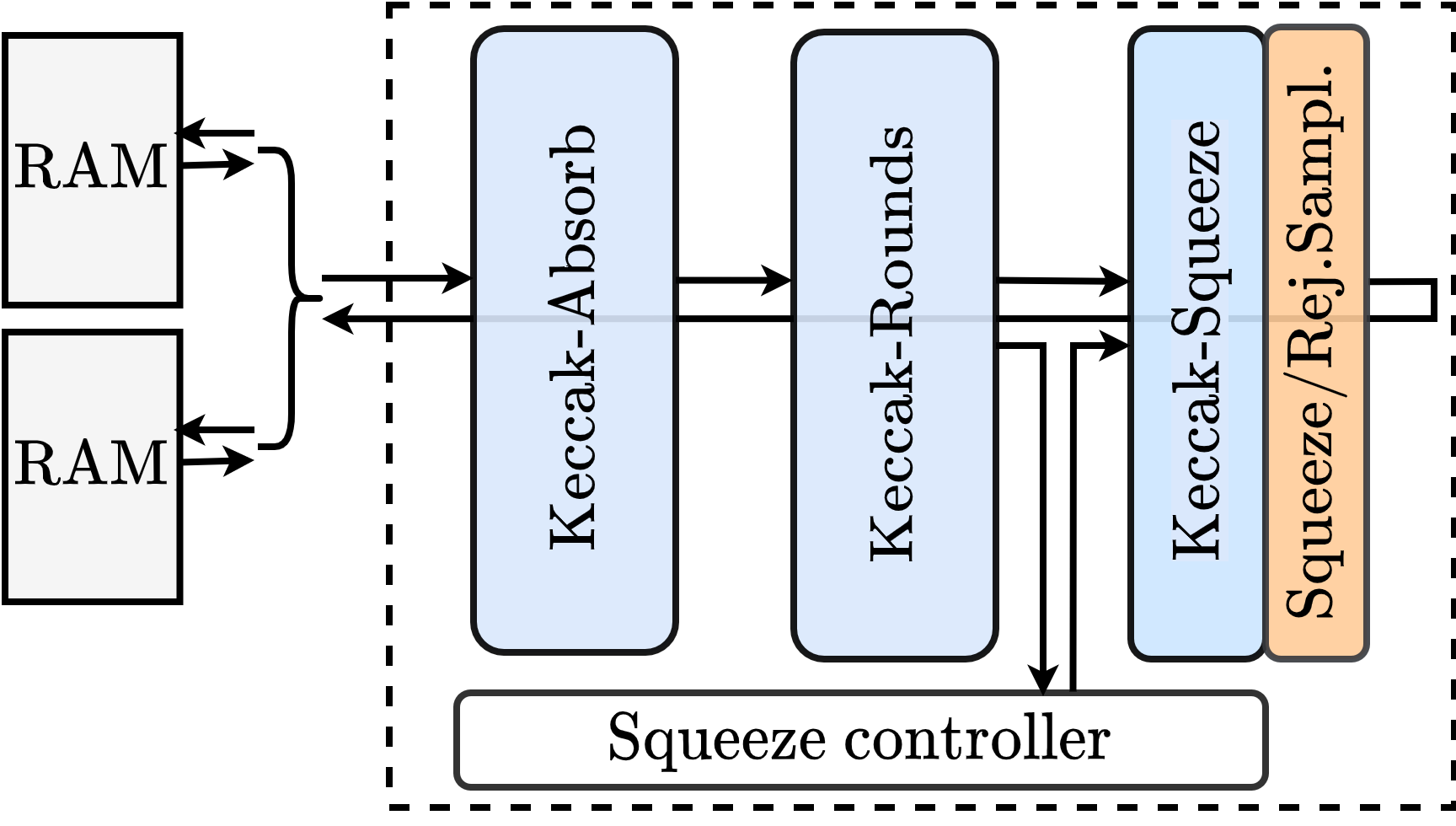}
    \vspace{-1em}
    \color{black}
    \caption{ High level architecture of Keccak that supports all the modes of Keccak as well as samplers}
    \color{black}
    \label{fig:keccak_detailed}
\end{figure}

\noindent\textbf{Other sampling operations:}
Saber uses a binomial sampler for generating secret polynomials which is implemented as a separate unit.
Dilithium requires three different rejection sampling: uniform, $\eta$, and $\gamma$ sampling. For the uniform and $\eta$ sampling, we need to extract 24 and 4 bits from the Keccak output buffer, respectively, and we can utilize Keccak output fully after every squeeze. On the other hand, the $\gamma$ sampling needs 18 or 20 bits from the Keccak output and it does not utilize the Keccak output fully after every squeeze with some leftover bits.

The same approach of shifting the output buffer and leftover bit buffer as described in the previous section can be used for $\gamma$ sampling as well. However, this leads to a Keccak output buffer generating six different types of outputs 4, 13, 18, 20, 24, and 64 bits. This can be controlled using a multiplexer, which in hardware is very expensive. In order to reduce the cost, we take an intermediate smaller buffer of size 192 bits (=$\mathrm{lcm}(4,24,64)$) and use it for squeezing the results for 4, 24, and 64 bits. The Keccak output buffer then outputs only four different types of outputs 13, 18, 20, and 192 bits, thus saving around $\approx$1200 LUTs in FPGAs. \textcolor{black}{ A high-level view of the complete architecture of SHA-SHAKE unit with samplers is shown in Fig.~\ref{fig:keccak_detailed}. The datapaths for Keccak modes and samplers are coalesced together into one unit. This is then controlled by the squeeze controller. }

\subsection{Remaining scheme-specific building blocks}
In the above sections, we discussed how we efficiently implement a common polynomial arithmetic unit for Dilithium and Saber. We also discussed the optimized implementation of other major area-consuming blocks. In this section, we discuss how we implement the remaining building blocks and various instructions that our cryptoprocessor provides to realize a flexible instruction-set cryptoprocessor. Note that the output given after the polynomial multiplication now has two coefficients per word instead of 4 coefficients per word storage style used in \cite{DBLP:journals/tches/RoyB20}. This causes Saber building blocks to consume more clock cycles however, we are able to avoid the extra clock cycles required for pre-processing the input for feeding into these modules.
Architectures for the Saber-specific modules are as follows.
\begin{itemize}
    \item Modules 'Unpack (Saber)', 'AddPack', and 'AddRound', 
     provided in Table~\ref{table:ins} Instruction Set-2, which can run in parallel with Keccak's SHA/SHAKE and all the other instructions provided in Instruction-Set 1.
    \item 'CMOV', 'Verify', and 'COPY'     also have the ability to run in parallel with instructions in Set-1. 
    \item Instruction 'Binomial Sampler' is provided in Set-1.  
    Thus, while one polynomial is being generated and passed through the binomial sampler, the previously generated polynomial can be transformed to the NTT domain in parallel. 
    \item Instruction 'BS2POLVEC' converts byte-stream to vector form as required by the NTT module for one polynomial at a time. and can be run in parallel with Set-2 instructions.
    
\end{itemize}

Architectures for the Dilithium-specific modules are as follows.
\begin{itemize}
    \item Module 'Pack-Unpack (Dilithium)', in Table~\ref{table:ins} Instruction-Set 1, allow unpacking/packing of polynomials while performing transformation or polynomial arithmetic on already/to-be unpacked/packed polynomials, in parallel.  
    \item Module 'SampleInBall' requires the memory to be refreshed with zeroes. For this 'Refresh Memory' instruction is provided in Set-1. Once the memory is refreshed, SampleInBall instruction provided in Set-2 can be used.
    \item Instruction 'Encode\_$\mathcal{H}$' provided in Set-2 is used to pack the polynomials which are then fed to SHAKE-256 for hashing.
    \item  Instructions 'Decompose' and 'Power2Round', in Set-1 and Set-2 respectively, are implemented as per the specification, consuming 128 clock cycles for processing one polynomial.
    \item 'MakeHint' and 'UseHint' modules require Decompose function. Since this is an instruction-set processor we use the existing Decompose module instead of making duplicate implementations and just implement the equality checkers in MakeHint and UseHint modules, which return the desired output.
    \item Instruction 'Counter\_ref' is provided in Set-2, to ensure that the MakeHint counter for the hamming weight becomes zero if the loop exit conditions are not satisfied.
    \item Dilithium uses the same seed with different nonce values for generating polynomials. Naively in hardware, we can make copies of seeds with different nonce values to generate different polynomials. However, this is inefficient and occupies extra memory. We provide a 'Write instruction' in Set-1 which can be used to store a nonce of user's choice after the seed. Thus, not only saves memory but allows the user flexibility to generate polynomials in any order. 
    \item Instruction 'Verify (Dilithium)' in Set-1 is used to verify the conditions for a valid signature during the signing procedure. In case the verification conditions are not met, it sends back the instruction pointer to the instruction which marks the start of the \textit{for} loop. 
\end{itemize}

\subsection{Parallel processing of instructions}

\begin{figure*}
    \centering
    \includegraphics[width=\textwidth]{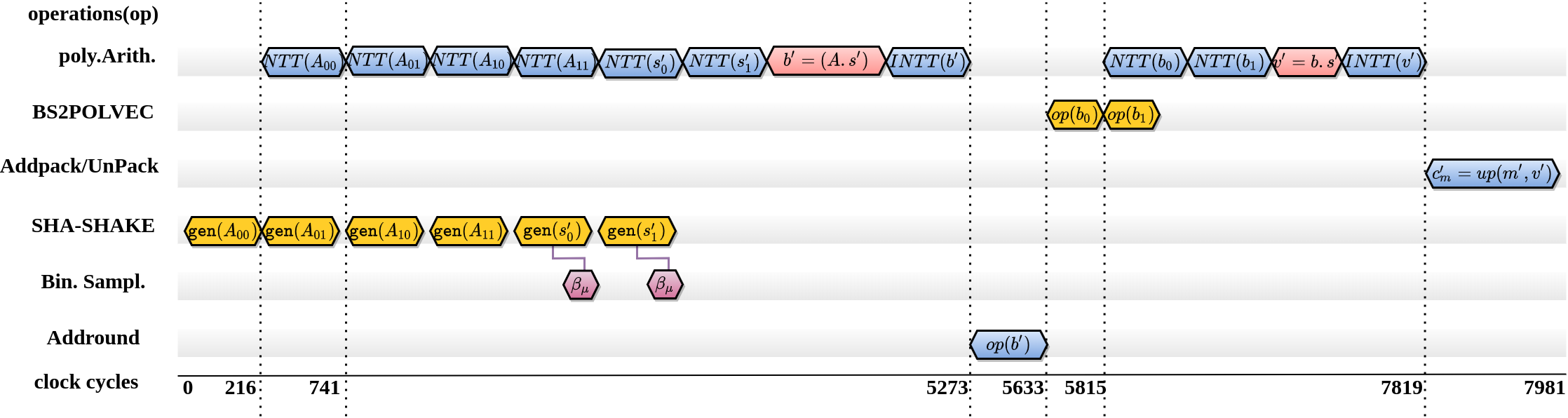}
    \vspace{-2em}
    \caption{Operation scheduling of LightSaber PKE encryption}
    \label{fig:ls_timeline2}
    \vspace{-1em}
\end{figure*}
We observe that both Dilithium and Saber protocols have data-independent instructions. Therefore, it is possible to speed up the two protocols by executing data-independent instructions in parallel (when possible). Such parallel processing of low-level operations has been explored in the works~\cite{NIST_presentation_Gaj, DBLP:journals/iacr/DangFAMNG20}. They design block-pipelined architectures (that are different from instruction-set architectures) and place the building blocks in such a way that several data-independent computation steps are overlapped (i.e., executed in parallel) using the dedicated blocks in the correct order. While an instruction-set-based cryptoprocessor offers significant flexibility over a block-pipelined (therefore specific) architecture, overlapping of data-independent `instructions' becomes a relatively challenging task. Implementing the parallel processing of instructions requires a complicated synchronization and control mechanism, e.g., scheduling of instructions, detecting completions of concurrent instructions, initiating new instructions, managing the data memory, etc. 
In our work, we apply overlapping of data-independent computations in the context of an \emph{instruction-set architecture} and execute data-independent Keccak-based and polynomial arithmetic-based operations in parallel. This strategy effectively reduces the overall cycle count at the cost of a negligible area overhead. 

When the instructions are run in parallel they need to read/write data from memory simultaneously. Since one BRAM can only offer us one set of read/write ports, we needed to come up with an efficient solution. To support the parallel execution of Keccak and polynomial arithmetic, we split the memory unit into four BRAM sets. While the NTT unit occupies read and write ports of any two BRAM sets, the Keccak unit works with the remaining two sets. This is discussed in detail later in Section \ref{sec:mem}. We also add a program controller unit that loads all the instructions in an instruction RAM and then sends them one by one to the compute core for processing in parallel or sequence as specified in the instruction. The two types of instructions are  stored together along with 4 control bits in the Instruction RAM. Fig.~\ref{fig:ls_timeline2} shows the instruction flow for LightSaber.\texttt{PKE.Enc()}. We can see the instructions in Set-1 working in parallel with instructions in Set-2 given in Table~\ref{table:ins}. We achieve a reduction of 10\%, 13\%,and 15\% in cycle count during decapsulation for LightSaber, Saber, and FireSaber respectively. For Dilithium signature generation we reduce the cycle count by 20\%, 25\%,and 28\% for Dilithium 2, 3, and 5 respectively.

 \begin{table}[!t]
  \caption{Instructions where an instruction from the first column can be run in parallel with an instruction from the second column.}
  \vspace{-1em}
 \footnotesize{
 \begin{tabular}{c l|l}
 \cline{2-3}
   & \textbf{Instruction Set-1}                                       & \textbf{Instruction Set-2} \\ \cline{2-3}
   
   \parbox[t]{0.2mm}{\multirow{5}{*}{\rotatebox[origin=c]{90}{S+D}}}
   & Reset Keccak                 & NTT    \\            
   & SHA-256/512                  & INTT \\
   & SHAKE-128/256                & Coefficient-wise multiply \\ 
   & SHAKE intermediate reset     & Coefficient-wise add \\           
   & SHAKE resume                 & Coefficient-wise subtract\\ \cline{2-3}

   \parbox[t]{0.2mm}{\multirow{3}{*}{\rotatebox[origin=c]{90}{Saber}}}                
   & SHAKE-128 (writing 26 bits)  & AddRound/AddPack  \\  
   & BS2POLVEC                    & Unpack (Saber)  \\
   & Binomial Sampler             & Verify/CMOV/COPY  \\
   \cline{2-3}
   
   \parbox[t]{0.2mm}{\multirow{6}{*}{\rotatebox[origin=c]{90}{Dilithium}}}
   & SHAKE-128 rej. $[0, q-1]$                                   & SampleInBall \\
   & SHAKE-256 rej. [-$\eta$, $\eta$]                            &  Encode\_$\mathcal{H}$ \\
   & SHAKE-256 rej. [-($\gamma$ - 1), $\gamma$ - 1]              &  Power2Round \\
   &  Pack-Unpack (Dilithium)                                    &  MakeHint \\
   & Decompose/Verify (Dilithium)                                &  UseHint \\     
   & Write Inst./Refresh Memory                                  &  Counter\_ref \\
   \cline{2-3}                   
 \end{tabular}}
 \label{table:ins}
 \end{table}

\subsection{Organization of data memory}\label{sec:mem}

We need to ensure that our cryptoprocessor has sufficient memory to support all the variants of both Dilithium and Saber. For this, we consider the highest security variants as they require matrices and vectors of the highest dimensions. The signature generation function in Dilithium and the decapsulation function in Saber are the most time-consuming functions and also require the highest amount of data storage. For the Dilithium variant with the NIST security level 5, the public matrix consists of 56 polynomials. During signing operation, we need to precompute and store the secret key consisting of 23 polynomials, in the NTT domain. FireSaber has much smaller requirements as its public matrix and secret vector comprise only 16 and the 4 polynomials respectively. Thus Dilithium-5 determines the overall memory requirement of the cryptoprocessor.

Storing the entire public matrix in the memory makes the signing operation faster as it is used in the for a loop several times. If we pre-compute and store all the polynomials, we require storage for 79 polynomials before the signing loop starts along with seeds and hash values. Also, we need to store intermediate results during the signing operation, which further increases the memory requirement. In the proposed work, we use the ability of our cryptoprocessor to process data-independent instructions in parallel. Instead of generating and storing the public matrix at once, we generate it on the fly in parallel to the polynomial multiplication operation, thus reducing more than half the memory requirement without compromising the performance.

As we discussed in the previous section our data memory length is 64-bit and so we store 2 coefficients per word of data memory. With all the constraints in consideration and flexibility requirements in place, the implementation of Saber requires only four BRAM36K units while Dilithium requires 20 BRAM36K. The proposed cryptoprocessor uses parallel memory organization to ensure efficient load and storage of polynomials. This is especially important for the parallel execution of NTT and Keccak operations. To that end, the memory was split across four major blocks, with each of them having five BRAM36K elements, which enables the parallel execution of NTT and Keccak. The constants for NTT and inverse NTT are kept in a ROM which is also interpreted using one BRAMs in our implementation. Along with this, the program controller which is used to load all instructions at once and then handle all the data-independent executions in parallel requires three BRAM36K.

\section{Results}\label{sec:results}

We present three different versions of cryptoprocessor implementation: $(i)$ the first implementation uses 23-bit Dilithium prime $2^{23} - 2^{13} + 1$ for both Dilithium and Saber's NTTs, $(ii)$ the second implementation uses 23-bit Dilithium prime and 24-bit prime $2^{24} - 2^{14} + 1$ for Dilithium and Saber's NTTs respectively, and $(iii)$ the third implementation uses 23-bit Dilithium prime and 25-bit prime $2^{25} - 2^{14} + 1$ for Dilithium and Saber's NTTs respectively. They give us comparative area consumption for the three choices. The first, second, and third implementations are referred to as 23-bit, 24-bit, and 25-bit implementations, respectively. The 23-bit and 24-bit implementations ensure negligible-error probability for multiplications in Saber while the 25-bit implementation performs error-free multiplication for Saber. All three implementations support all the subroutines and all the security levels of Dilithium and Saber using an instruction-set architecture framework. 

The proposed unified cryptoprocessor architecture is described entirely in Verilog and it is implemented for FPGA and ASIC platforms. For FPGA, the 23-bit, 24-bit, and 25-bit implementations are synthesized and implemented using Vivado 2019.1 for the target platform Zynq Ultrascale+ ZCU102 with an area-optimized implementation strategy. The FPGA implementations achieve a 200 MHz clock frequency. In Table~\ref{tab:utilize}, we present the resource utilization of the proposed cryptoprocessor and its main building blocks. While the multiplier and Keccak (with sampler) blocks are able to achieve a maximum frequency of 260MHz, the remaining building blocks can operate at higher frequency. The implementations with 23-bit/24-bit/25-bit prime use 18,490/18,494/18,406 LUTs (6.9\%), 9,308/9,319/9,323 DFFs (1.7\%), 4 DSPs (0.1\%) and 24 BRAMs (2.6\%). The number of BRAMs in our cryptoprocessor is determined by the memory requirement of Dilithium since it is significantly more memory-consuming than Saber. 
The Keccak and multiplier units together consume more than half of the overall area. For ASIC, the 24-bit implementation is synthesized with UMC 65nm library with low leakage (LL) process device option using Cadence tool. For the ASIC implementation, DSP blocks used for integer multiplications in FPGA are replaced with ASIC multipliers. The rest of the arithmetic logic remains the same. The ASIC implementation synthesis achieves 400 MHz clock frequency with 0.317mm$^2$ area ($\approx$220 kGE) excluding on-chip memory.
\color{black}

Table~\ref{tab:time} presents the cycle count and latency (in $\mu$s) for the operations of Dilithium and Saber for different security levels in FPGA. With 200 MHz clock frequency in FPGA, the CCA-secure key generation, encapsulation, and decapsulation operations for Saber take 54.9, 69.7, and 94.9 $\mu$s, respectively. We divide Dilithium's signature generation into three parts, pre-sign, sign, and post-sign, and report their performances separately. For a signature generation, the pre-sign and post-sign parts are performed only once while the sign part is repeated until a valid signature is generated. We report the performance for the \textit{best-case scenario} where the valid signature is generated after the first loop iteration. The key generation, signing and verification operations for Dilithium-3 take 114.7, 237, and 127.6 $\mu$s, respectively, in the FPGA. In ASIC the speed of the cryptoprocessor improves by 2$\times$ as the clock frequency increases by 2$\times$.

\begin{table}
\color{black}
    \caption{Area of the cryptoprocessor on the Zynq Ultrascale+ ZCU102 FPGA platform. All security levels of Dilithium and Saber are supported.} %
    \label{tab:utilize}
    \vspace{-1em}
    \centering
    \begin{tabular}{l l|c|c|c|c}
    
    \cline{2-6}
       & \textbf{Unit}  & \textbf{LUTs} & \textbf{FFs} & \textbf{DSPs} & \textbf{BRAMs}  \\ \cline{2-6}
       & \textbf{ComputeCore} & \textbf{17,164} & \textbf{9,027} & \textbf{4} & \textbf{21} \\ \cline{2-6}
       \parbox[t]{0.2mm}{\multirow{8}{*}{\rotatebox[origin=c]{90}{Saber}}} 
       & $\lfloor$AddPack & 398 & 548 & 0 & 0 \\
       & $\lfloor$AddRound & 359 & 362 & 0 & 0\\
       & $\lfloor$BS2POLVEC & 340 & 360 & 0 & 0\\
       & $\lfloor$Unpack (Saber) & 393 & 406 & 0 & 0\\
       & $\lfloor$Verify (Saber) & 103 & 208 & 0 & 0\\
       & $\lfloor$CMOV & 12 & 34 &0&0\\
       & $\lfloor$COPY & 28 & 98 & 0&0\\
       & $\lfloor$Sampler & 934 & 710 & 0 & 0\\ \cline{2-6}
       \parbox[t]{0.2mm}{\multirow{10}{*}{\rotatebox[origin=c]{90}{Dilithium}}} 
       & $\lfloor$Decompose & 548 & 286 & 0 & 0\\
       & $\lfloor$Power2Round & 160 & 64 & 0 & 0\\
       & $\lfloor$MakeHint & 245 & 119 & 0 & 0\\
       & $\lfloor$UseHint & 609 & 393 & 0 & 0\\
       & $\lfloor$Encode\_$\mathcal{H}$ & 122 & 231 & 0 & 0\\
       & $\lfloor$Pack (Dilithium) & 740 & 151 & 0& 0\\
       & $\lfloor$Unpack (Dilithium) & 296 & 156 & 0& 0\\
       & $\lfloor$SampleInBall & 485 & 260 & 0 & 0\\
       & $\lfloor$Refresh & 4 & 7 & 0 & 0 \\
       & $\lfloor$Verify (Dilithium) & 28 & 69 & 0 & 0\\ \cline{2-6}
       \parbox[t]{0.2mm}{\multirow{5}{*}{\rotatebox[origin=c]{90}{S+D}}} 
       & $\lfloor$Memory & 268 & 12 & 0 & 20 \\
       & $\lfloor$Keccak+Sampler & 8,738 & 3,482 & 0 & 0 \\
       & $\lfloor$ Multiplier$^a$ &  2,341& 1,048 & 4 & 1      \\
                                 & $\lfloor$ Multiplier$^b$ & 2,347 & 1,059 & 4 & 1  \\
                                  & $\lfloor$ Multiplier$^c$ & 2,257 & 1,079 & 4 & 1\\
       \cline{2-6}
       
       & \textbf{ProgramController} & \textbf{1,371} & \textbf{260} & \textbf{0} & \textbf{3} \\\cline{2-6}
       & \textbf{Total} & \textbf{18,494} & \textbf{9,319} & \textbf{4} & \textbf{24} \\ \cline{2-6}
    
    \multicolumn{6}{l}{%
    \begin{minipage}{8.0cm}%
    $^a$:for 23-bit implementation. \\
    $^b$:for 24-bit implementation. (used for total computation) \\
    $^c$:for 25-bit implementation.
    \end{minipage}%
    }\\
    
    \end{tabular}
    \color{black}
\end{table}

\begin{table}[t!]
\footnotesize{
    \caption{Performance results for Saber-KEM and Dilithium in FPGA}
    \label{tab:time}
    \vspace{-1em}
    \centering
    \begin{tabular}{l|r|r|r|r|r|r} 
    \hline
        \multirow{3}{*}{\textbf{Operation}}   & \multicolumn{2}{c|}{\textbf{LightSaber}} & \multicolumn{2}{c|}{\textbf{Saber}} & \multicolumn{2}{c}{\textbf{FireSaber}} \\ 
        & \multicolumn{2}{c|}{\textbf{Dilithium-2}} & \multicolumn{2}{c|}{\textbf{Dilithium-3}} & \multicolumn{2}{c}{\textbf{Dilithium-5}} \\ \cline{2-7}
                                              & \textbf{Cycle} & \textbf{Lat.}$^\star$ & \textbf{Cycle} & \textbf{Lat.}$^\star$ & \textbf{Cycle} & \textbf{Lat.}$^\star$ \\ \hline
        Sab.Keygen        & 5,935  &  29.6 & 10,980 &  54.9 & 17,523 &  87.6 \\
        Sab.Encaps        & 8,081  &  40.4 & 13,941 &  69.7 & 21,603 & 108.0 \\
        Sab.Decaps        & 11,678 &  58.3 & 18,991 &  94.9 & 27,890 & 139.4 \\ \hline 
        Dil.Gen           & 14,183 &  70.9 & 22,957 & 114.7 & 38,841 & 194.2 \\
        Dil.Sign$_{pre}$  & 7,554  &  37.7 &  9,273 &  46.3 & 12,448 &  62.2 \\
        Dil.Sign   & 21,115 & 105.5 & 35,865 & 179.3 & 52,955 & 264.7 \\
        Dil.Sign$_{post}$ & 1,689  &   8.4 &  2,280 &  11.4 &  3,057 &  15.2 \\
        Dil.Verify        & 15,044 &  75.2 & 25,535 & 127.6 & 45,789 & 228.9 \\
      \hline
    \multicolumn{7}{l}{%
    \begin{minipage}{8.0cm}%
   $^\star$: Latency for FPGA implementation in $\mu s$.
    \end{minipage}%
    }\\
    \end{tabular}
    }
\end{table}


\subsection{Comparison with the existing results}
\begin{figure}
\color{black}
    \centering
    \includegraphics[width=0.45\textwidth]{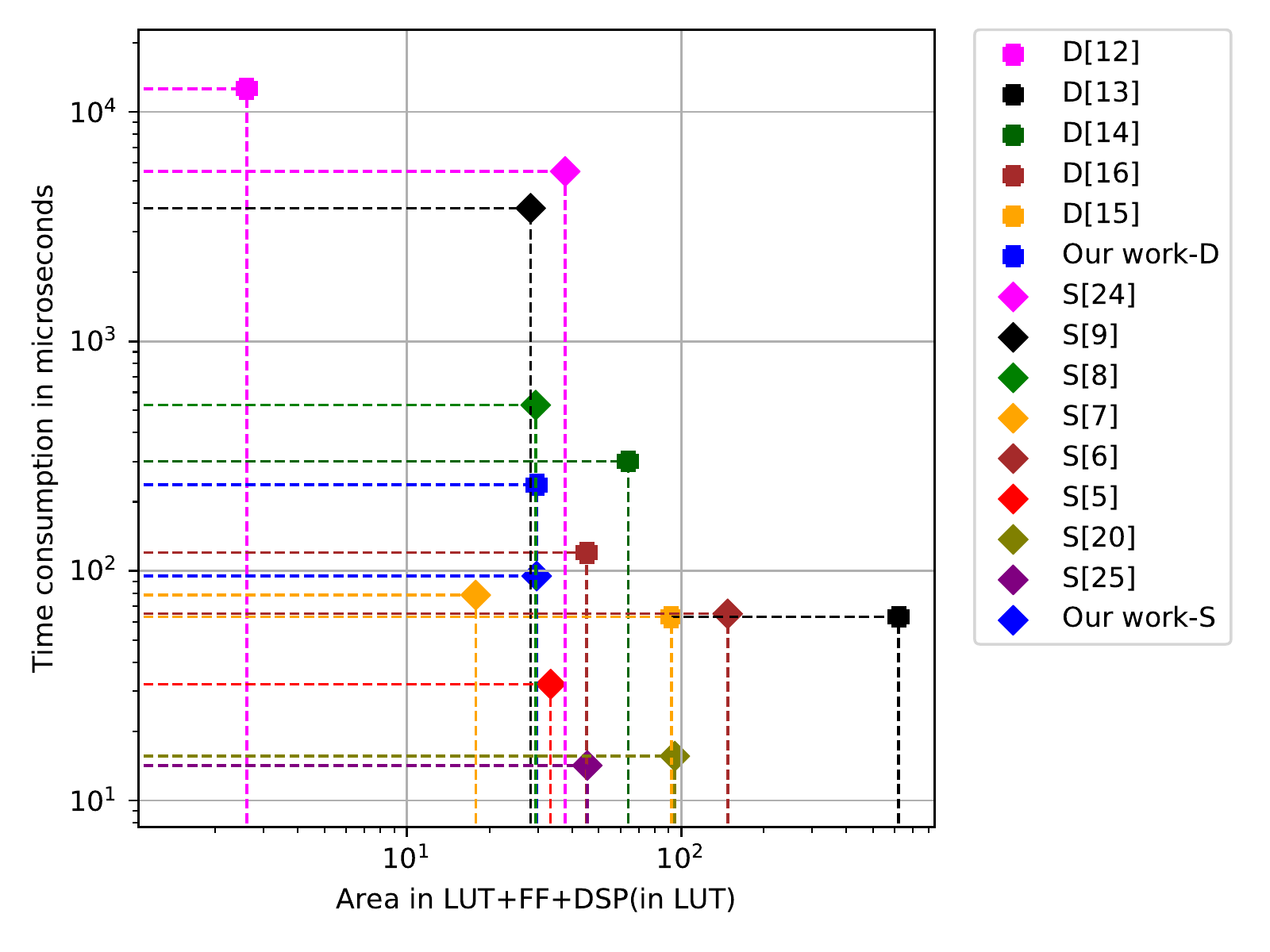}
    \caption{ The figure presents a graphical representation of time-consumption and resource comparisons between our work and existing works. The lower the time-consumption and resource consumption the better. Logarithmic scale is used for both axes. With both the schemes included, our cryptoprocessor still consumes reasonable resources and provides optimal performance. Each DSP unit is assumed to consume 483 LUTs (obtained with Xilinx Vivado IP Generator).}
    \label{fig:comp_graph}
\end{figure}

We provide comparisons of our cryptoprocessor with related works in the literature in terms of area, performance, and flexibility for Dilithium-3 and Saber as shown in Table~\ref{tab:comp_dil} and Table~\ref{tab:comp_saber}, respectively. A graphical representation of the performance and resource consumption is given in Fig.~\ref{fig:comp_graph}. For simplicity and better understanding, we divide the comparisons into three parts: $(i)$ comparisons with other designs supporting multiple schemes, $(ii)$ comparisons with standalone implementations of Dilithium, and $(iii)$ comparisons with standalone implementations of Saber.
 
\noindent\textbf{Comparisons with unified architectures:} 
We note that only a few works target unified architectures that support multiple PQC schemes~\cite{sapphire_2019,risqv_2020,fritzmann_masked_21,DangMG21}. Their area and performance results along with our architecture are presented in Table~\ref{tab:comp_saber}. In~\cite{sapphire_2019}, the authors present \textit{Sapphire}, a cryptoprocessor coupled with RISC-V processor implemented in ASIC for various lattice-based Round 2 schemes in NIST's PQC standardization. It does not support Saber, while the results provided for Dilithium use the outdated Round-2 specifications. In~\cite{risqv_2020}, the authors present a RISC-V architecture tightly coupled with hardware accelerator for Crystals-Kyber, NewHope, and Saber. Compared to the Saber implementation in~\cite{risqv_2020}, our FPGA and ASIC implementations show up to 304$\times$ and 564$\times$ better performances, respectively. The work in~\cite{fritzmann_masked_21} presents a HW/SW co-design of Crystals-Kyber and Saber. Our implementation shows superior performance in terms of both speed and area as we target an implementation entirely in hardware. \textcolor{black}{The authors in \cite{DangMG21} present high-speed design results for Kyber, NTRU, and Saber. Our unified architecture consumes much less area compared to their Saber results.}

\noindent\textbf{Comparisons with Dilithium-only implementations:}
There are only a few FPGA-based implementations of Dilithium~\cite{zhou_dil_21,ricci_dil_21,DBLP:journals/iacr/LandSG21,gaj_dil_21,dilithium_2022} in the literature. Their area and performance results along with our work are presented in Table~\ref{tab:comp_dil}. Similar to 'Saber only architecture', these architectures have an obvious limitation of providing support for only digital signature and not PKE/KEM. Zhou~\textit{et al.}~\cite{zhou_dil_21} propose a HW/SW co-design by offloading computationally intensive operations such as SHA3/SHAKE and polynomial multiplication to the hardware while keeping the rest of the operation in the software. Their implementation has a small area as they implement only a few building blocks in hardware. Our pure-hardware solution shows almost up to two orders of magnitude better performance compared to their HW/SW co-design solution. 

In~\cite{ricci_dil_21}, the authors present three high-performance architectures for Dilithium targeting FPGA. Their implementations can perform  key generation, signature generation, and verification operations in 51.9, 63.1, and 95.1 $\mu$s, respectively. Although they show better performance than our implementation, their implementation for sign operation- consumes $3.5\times$, $9.2\times$, $241.2\times$, and $6\times$ more LUTs, DFFs, DSPs, and BRAMs compared to our implementation. Moreover, our work can perform all three operations in a single implementation. The FPGA implementation of Dilithium~\cite{DBLP:journals/iacr/LandSG21} targets reducing LUT utilization by employing extra DSP units for computations. Their implementation utilizes $1.5\times$ more LUTs and $11.2\times$ more DSPs units. Our work shows $1.47\times$, $1.08\times$, and $1.36\times$ better performance for the key generation, signature generation, and signature verification operations, respectively. In~\cite{gaj_dil_21}, a high-performance Dilithium implementation is presented. It shows better performance at the expense of $3\times$ and $4\times$ more LUT/DFF and DSP, respectively. Similarly, in~\cite{dilithium_2022} the authors achieve better performance at the expense of area. Our implementation is $2\times$ slower but consumes $1.5\times$ less area and provides the flexibility to do the operations in parallel or sequentially.

\noindent\textbf{Comparisons with Saber-only implementations:} 
There are several works in the literature implementing Saber in hardware, e.g.,~\cite{DBLP:journals/tches/RoyB20,gaj_fpt_19,hecompact,abdulgadirlightweight,mera_compact_20,fritzmann_masked_21} on FPGA and~\cite{risqv_2020,imran2021design,lwrpro_21,GhoshMKDGVS22} on ASIC platforms. Their area and performance results along with our work are presented in Table~\ref{tab:comp_saber}. Their obvious limitation is that they only provide support for PKE/KEM but not a digital signature scheme. As we discussed in Section~\ref{sec:mem}, the area of our design is determined by the Dilithium scheme, therefore it is expected to consume more area than Saber-only implementations.  \textcolor{black}{However, when we compare our design to high-performance implementations of Saber, we consume much less area compared to~\cite{DBLP:journals/tches/RoyB20,abs}, and almost a similar area compared to ~\cite{imran2021design,lwrpro_21}. We consume 2$\times$ more area compared to \cite{GhoshMKDGVS22} and deliver 3.5$\times$ better performance. }Our unified cryptoprocessor outperforms~\cite{abdulgadirlightweight,mera_compact_20} and shows a similar performance compared to the architectures in~\cite{gaj_fpt_19,hecompact}.

\begin{table}
\scriptsize{
    \caption{Comparison Table for Dilithium-3}
    \label{tab:comp_dil}
    \vspace{-1em}
    \centering
    \begin{tabular}{l|l|l|l|l}
    \hline
        \multirow{2}{*}{\textbf{Ref.}} &  \multirow{2}{*}{\textbf{Plat.}} & \textbf{Performance} & \textbf{Freq.} & \textbf{Area} (\textit{mm}$^2$ or LUT/ \\
                                       &                                   &  (in $\mu$s)         & (\textit{MHz}) &
                                       FF/DSP/BRAM) \\ \hline
       
       \cite{zhou_dil_21}$^{\dag}$        & Zynq                     & -/12.6K/9.9K   & 100  &  2.6K/-/-/-       \\\hline
       
       \cite{ricci_dil_21}$^a$            & \multirow{3}{*}{US+} & 51.9/-/-         & 350  &  54.1K/25.2K/182/15  \\
       \cite{ricci_dil_21}$^{b,d}$        &                      & -/63.1/-         & 333  &  68.4K/86.2K/965/145 \\
       \cite{ricci_dil_21}$^c$            &                      & -/-/95.1         & 158  &  61.7K/34.9K/316/18  \\\hline
       
       \cite{DBLP:journals/iacr/LandSG21}$^{d}$ & \multirow{2}{*}{Ar.-7} & 229/0.3k/0.2k & \multirow{2}{*}{145} &  \multirow{2}{*}{30.9K/11.3K/45/21}  \\
       \cite{DBLP:journals/iacr/LandSG21}$^{e}$ &                         & 229/0.85k/0.2k &                      &                                      \\\hline
        \cite{dilithium_2022}$^{d}$ & \multirow{2}{*}{Ar.-7} & 60/0.12k/63.8 & \multirow{2}{*}{96.9} &  \multirow{2}{*}{30K/10.34K/10/11}  \\
       \cite{dilithium_2022}$^{e}$ &                         & 60/0.46k/63.8 &                      &                                      \\\hline
       
       \cite{gaj_dil_21}$^{d}$ & \multirow{2}{*}{US+} & 32/63/39 & \multirow{2}{*}{145} &  \multirow{2}{*}{55.9K/28.4K/16/29}  \\
       \cite{gaj_dil_21}$^{e}$ &                         & 32/193/39 &                      &                                      \\\hline
       
       \multirow{2}{*}{\textbf{Our}$^{d,f}$} & US+    & 114.7/237/127.6           &  200  &  18.5K/9.3K/4/24      \\
                                            & 65nm  & {\tiny{$\approx$}}57.4/118.5/63.8  &  400  &  $\approx$0.317+1.230 \textit{mm}$^2$  \\\hline

    \multicolumn{5}{l}{
    \begin{minipage}{8.0cm}
    $^a$: Works for K.Gen.
    $^b$: Works for Sign.
    $^c$: Works for Verify. \\
    $^d$: Reports best-case scenario. 
    $^e$: Reports average-case scenario. \\
    $^f$: Supports multiple schemes. 
    $^{\dag}$: HW/SW co-design.
    \end{minipage}
    }\\
    \end{tabular}
    }
\end{table}

\begin{table}
\scriptsize{
    \caption{Comparison Table for Saber-KEM}
    \label{tab:comp_saber}
    \vspace{-1em}
    \centering
    \begin{tabular}{l|l|l|l|l}
    \hline
        \multirow{2}{*}{\textbf{Ref.}} & \multirow{2}{*}{\textbf{Plat.}} & \textbf{Performance} & \textbf{Freq.}          & \textbf{Area} (\textit{mm}$^2$ or LUT/\\
                                       &                                    & (in $\mu$s)       & (\textit{MHz})  & FF/DSP/BRAM)          \\ \hline
       
       \cite{risqv_2020}$^{b}$      & 65nm   &  16K/21K/26K  & 45.47 &  0.914 \textit{mm}$^2$  \\\hline
       \cite{imran2021design} & 65nm   &  7.1/7.1/9.3        & 1000  &  0.314 \textit{mm}$^2$  \\\hline
       \cite{lwrpro_21}       & 40nm   &  2.7/3.6/4.3        & 400   &  0.380 \textit{mm}$^2$  \\\hline
       \color{black}
       \cite{GhoshMKDGVS22}& 65nm   &  89,6/116.9/146.18       & 40-160   &  0.158 \textit{mm}$^2$  \\\hline
       \color{black}

       \cite{fritzmann_masked_21}$^{\dag,b}$ & Ar.-7 & 3.6K/4.9K/5.5K  & 62.5 &  20K/11K/13/36.5  \\\hline
       \cite{mera_compact_20}$^{\dag}$     & Ar.-7 & 3.2K/4.1K/3.8K  & 125  &  7.4K/7.3K/28/2       \\\hline
       \cite{abdulgadirlightweight}        & Ar.-7 & –/467.1/527.6   & 100  &  6.7K/7.3K/32/0       \\\hline
       \cite{hecompact}                    & US+     & 48.9/63.2/78.5  & 250  &  10.1K/7.7K/0/3       \\\hline
       \cite{gaj_fpt_19}$^{\dag}$          & US+     & -/60/65         & 322  &  12.5K/11.6K/256/4    \\\hline
       \cite{DBLP:journals/tches/RoyB20}   & US+     & 21.8/26.5/32.1  & 250  &  23.6K/9.8K/0/2       \\\hline
       \color{black}
       \cite{abs}  & US+     & 10.2/12.6/15.6  & 250  &  41.5K/22.3K/64/2       \\\hline
       \color{black}
       \cite{DangMG21} & US+     & 7.6/10.5/14.2 & 400  &  21.6K/12.25K/24/10.5    \\\hline
       \color{black}
       
       \multirow{2}{*}{\textbf{Our}$^{a,b}$} & US+    & 54.9/69.7/94.9           &  200  &  18.5K/9.3K/4/24      \\
                                            & 65nm  & $\approx$27.5/34.9/47.5  &  400  &  $\approx$0.317+1.230 \textit{mm}$^2$  \\\hline
                                      
    \multicolumn{5}{l}{%
    \begin{minipage}{8.0cm}%
    $^a$:On-chip memory area is estimated as $\approx$1.230 \textit{mm}$^2$. \\
    $^b$:Supports multiple schemes.
    $^{\dag}$: HW/SW co-design.
    \end{minipage}%
    }\\
    \end{tabular}
    }
\end{table}

\label{sec:main}

\section{Discussions} \label{sec:disc}
In this work, we focused on Dilithium and Saber as a case study. As the proposed cryptoprocessor is an instruction set architecture, it can  be extended to provide support for various other lattice-based schemes.

\textcolor{black}{
For example, the support for the lattice-based scheme CRYSTALS-Kyber~\cite{kyber_nist_round3} can be added to our cryptoprocessor. Kyber has an NTT-friendly prime modulus and uses Keccak-based hash functions and pseudo-random number generation, which are the major time- and area-consuming building blocks. Our architecture already has building blocks for NTT and Keccak-based operations, which can be utilized using 'instructions'. However, Kyber uses an incomplete NTT and a slightly different algorithm for point-wise multiplication operation. Therefore, our cryptoprocessor can support Kyber by modifying the NTT unit and integrating Kyber-specific building blocks. }

In this work, our goal was to design an efficient and compact cryptoprocessor for Dilithium and Saber. A knowledge of compatibility between different schemes is very important to use  in real-life applications. However, the security of such implementations is also a valid concern. Several works exist in the literature for masking Saber~\cite{mask_saber} and Dilithium~\cite{mask_dil}, individually. We note that the traditional masking schemes are sufficient for masking the common building blocks of our compact cryptoprocessor.  The Keccak block can be efficiently secured against side-channel analysis using a boolean masking scheme~\cite{keccak_mask}. For the polynomial arithmetic operations, arithmetic masking is required. We can either use the same unit for operations on all multiple shares, saving area but consuming more clock cycles, or instantiate multiple times for each share, saving clock cycles but consuming more area. 

\section{Conclusion}
\label{sec:conclusion}
By designing a unified hardware architecture for the two finalists \dilname and Saber KEM of the NIST Post Quantum Cryptography Standardization, we showed that it is possible to realize a compact yet fast cryptoprocessor for performing both post-quantum key exchange and digital signature on ASIC and FPGA platforms. 

The optimized cryptoprocessor architecture greatly benefits from the algorithmic and structural similarities in the two implemented cryptographic schemes. The most expensive operations in both Dilithium and Saber are polynomial multiplications, and Keccak-based SHA3 and SHAKE computations. We demonstrated that by instantiating a unified NTT-based polynomial multiplier, we can compute the polynomial multiplications of both schemes. Furthermore, by using a special prime modulus for computing the NTTs of Saber, we can greatly minimize the area overhead of the unified multiplier compared to a Dilithium-only multiplier. Similarly, starting from a high-speed Keccak core, we designed an optimized wrapper around it to pre-process the inputs and post-process the outputs of SHA3 and SHAKE on the fly, and by doing so we effectively reduced the number of unnecessary memory read and write cycles. 

Finally, with all the optimizations, our unified cryptoprocessor on a Xilinx FPGA computes Saber's key generation, encapsulation, and decapsulation in 54.9, 69.7, and 94.9 $\mu$s respectively; and Dilithium-3's  key generation, signing (best case), and verification in 114.7, 237, and 127.6 $\mu$s respectively. The designed cryptoprocessor is even faster or smaller than several of the previously published works on Dilithium-only implementations on hardware platforms.

In the future, we intend to integrate more lattice-based schemes while keeping the design lightweight. We also intend to design and implement unified countermeasures for protecting our cryptoprocessor from side-channel and fault attacks in low time and area overheads.

\section*{Acknowledgments}
This work was supported in part by the Semiconductor Research Corporation through SRC task 3043.001, and by the State Government of Styria, Austria -- Department Zukunftsfonds Steiermark.

\ifCLASSOPTIONcaptionsoff
  \newpage
\fi

\bibliographystyle{IEEEtran}
\bibliography{sample-base}
\vspace{-1.5cm}
\begin{IEEEbiographynophoto}{Aikata Aikata} is a PhD student at Institute of Applied Information Processing and Communications, Graz University of Technology. Her research interests include  lattice-based cryptography and HW design. 
\end{IEEEbiographynophoto}
\vspace{-1.5cm}
\begin{IEEEbiographynophoto}{Ahmet Can Mert}
 is a postdoctoral researcher at the Institute of Applied Information Processing and Communications, Graz University of Technology, Austria. His research interest include homomorphic encryption, lattice-based cryptography and  HW design. 
\end{IEEEbiographynophoto}
\vspace{-1.5cm}
\begin{IEEEbiographynophoto}{David Jacquemin}
 is a PhD student at Institute of Applied Information Processing and Communications, Graz University of Technology. His research interests are isogeny, lattice-based cryptography.
\end{IEEEbiographynophoto}
\vspace{-1.5cm}
\begin{IEEEbiographynophoto}{Amitabh Das}  is a Principal Member of Technical Staff in the area of HW security architecture in the Product Security Office/Security R\&D group at AMD, Austin, Texas, USA. His research interests include HW cryptography, physical SCA and countermeasures, ML security, and SoC \& IP HW security. 
\end{IEEEbiographynophoto}
\vspace{-1.5cm}
\begin{IEEEbiographynophoto}{Donald Matthews} works at AMD in the Product Security Organization on a variety of projects including leading the AMD PQC effort. 
\end{IEEEbiographynophoto}
\vspace{-1.5cm}
\begin{IEEEbiographynophoto}{Santosh Ghosh} is a Research Scientist in Intel Labs. His research interests include LWC and PQC algorithms, low overhead memory safety architecture using LWC to solve long-lasting SW bugs \& vulnerabilities and to resist SCA, cryptographic HW microarchitecture and RTL, investigate and develop timing, power, EM and Photon SCA countermeasures.
\end{IEEEbiographynophoto}
\vspace{-1.5cm}
\begin{IEEEbiographynophoto}{Sujoy Sinha Roy} is an Assistant Professor of cryptographic engineering at IAIK, Graz University of Technology. He is a Co-Designer of “Saber”, which was a finalist KEM candidate in NIST’s PQC Standardization Project. He is interested in the implementation aspects of cryptography.
\end{IEEEbiographynophoto}

\end{document}